\documentclass[twocolumn,nofootinbib,aps,prd]{revtex4-2}
\pdfoutput=1 
\usepackage{indentfirst}
\usepackage[mathcal]{euscript}
\usepackage{amsfonts}
\usepackage{mathrsfs}
\usepackage{amsmath}
\usepackage{amssymb}
\usepackage{xcolor}
\usepackage{mathtools}
\usepackage{tensor}
\usepackage{physics}
\usepackage{graphicx}
\usepackage{bm}
\usepackage{upgreek}
\usepackage{braket}
\usepackage{color,soul}
\usepackage{csquotes}
\usepackage[small]{caption}
\usepackage{subcaption}
\usepackage{orcidlink}
\usepackage[normalem]{ulem}
\usepackage{yfonts}
\hypersetup{
            colorlinks,
            linkcolor=[rgb]{0,0.3,0.6}, 
            citecolor=[rgb]{0,0.3,0.6}, 
            urlcolor=[rgb]{0,0.3,0.6}
           }

\def\be{\begin{eqnarray}}
\def\ee{\end{eqnarray}}

\begin{document}

\title{Quasinormal modes of a quantum inspired black hole in four dimensions with cosmological constant}

\author{R. D. B. Fontana  \orcidlink{0000-0002-8835-7447}}\email{rodrigo.dalbosco@ufrgs.br}
\affiliation{Programa de Pós-Graduação em Física, Universidade Federal do Rio Grande do Sul - UFRGS, 91501-970 Porto Alegre RS, Brazil}
 \affiliation{Campus Tramandaí-RS Estrada Tramandaí-Osório, Tramandaí, UFRGS, CEP 95590-000, RS, Brazil.}

 \author{Jeferson de Oliveira \orcidlink{0000-0002-0874-3613}}\email{jeferson.oliveira@ufmt.br}
 \affiliation{ Instituto de Física, Universidade Federal de Mato Grosso, Cuiabá, Mato Grosso, 8060-900, Brazil}
 
 \author{B. Eslam Panah  \orcidlink{0000-0002-1447-3760}}\email{eslampanah@umz.ac.ir}
\affiliation{Department of Theoretical Physics, Faculty of Science, University of Mazandaran, P. O. Box 47416-95447, Babolsar, Iran}
\affiliation{Center for Theoretical Physics, Khazar University,
41 Mehseti Str., Baku, AZ1096, Azerbaijan.}
 
 \author{Angel Rincon \orcidlink{0000-0001-8069-9162}}\email{angel.rincon@physics.slu.cz}
\affiliation{ Departamento de Física, Universidad del Bío-Bío, Casilla 5-C, Concepción, Chile}
\affiliation{Research Centre for Theoretical Physics and Astrophysics, Institute of Physics, Silesian University in Opava,
Bezručovo náměstí 13, CZ-74601 Opava, Czech Republic}
\affiliation{Instituto Universitario de Matemática Pura y Aplicada, Universitat Polit\`ecnica de Val\`encia, Valencia 46022, Spain}


\date{\today}

\begin{abstract}
We study the scalar and Dirac perturbations of quantum-corrected black holes with cosmological constant. 
Using two different methods (WKB and double-null characteristic integration) we compute the quasinormal modes (QNMs) of de Sitter and Anti-de Sitter solutions considering linear field perturbations in the background geometry. In the  limit of general relativity black holes our methods demonstrate good convergence with the available results in literature. In the presence of an extra quantum parameter we verify that all perturbations are stable evolving in towers of quasinormal oscillations. We scrutinize the spectra of both dS and AdS solutions studying the influence of that extra parameter in the frequencies.

\end{abstract}



\maketitle

\section{Introduction}
\label{sec:intr}

Black holes are among the most rigorously confirmed solutions to Einstein’s field equations, representing fundamental objects in the theoretical framework of general relativity (GR). That solutions provide a complete description of the spacetime and its dynamical properties. The study of black hole physics is essential for the understanding of the fundamental laws of nature. As extreme objects, they probe the limits of classical and quantum aspects of gravitation, offering an unique arena to investigate their interplay. Owing to the presence of an event horizons that divide spacetime into causally disconnected regions, they provide deep insights into the nature of spacetime, thermodynamics, and quantum information theory, with far-reaching implications for the unification of gravity and quantum mechanics. Black holes can still be considered as laboratories for testing the predictions of alternative theories of gravity, offering valuable insights into potential deviations from GR. Research on the topic provided fundamental insights into the modern understanding of gravitation and quantum mechanics, most notably through the prediction of Hawking radiation, which establishes a deep connection between quantum mechanics and thermodynamics, thereby underscoring the quantum nature of spacetime.

By incorporating quantum effects, Hawking showed in the 1970s that black holes emit thermal radiation similar to that of a black body \cite{Hawking:1974rv,Hawking:1975vcx}. At the same time, Bekenstein introduced the idea of a black hole entropy $S$, proportional to the area of its event horizon $A_H$ \cite{Bekenstein:1973ur,Bekenstein:1974jk}. Both proposals led to remarkable new interpretations of the thermodynamic quantity, in contrast with the conventional notion that entropy is usually proportional to the volume of a body or system. These works revealed that black holes are not only geometric objects, but also thermodynamical systems. Thus, for a complete understanding of the black hole physics, the combination of thermodynamics, gravity, and quantum mechanics is essential \cite{Wald:1999vt,Carlip:2014pma,Carlip:2001wq,Nozari:2006ka}.

Combining the latter ingredients is particularly challenging, since the interactions among their components generate non-trivial effects in the black hole physics. Of particular interest is the fact that the heat capacity of a black hole is negative in many relevant cases, which means that the geometry is unstable and its temperature rises during particle emission. This leads to extreme scenarios such as infinite temperatures and singularities at the end of the evaporation process. These problems highlight the limitations of treating black holes solely as thermodynamical systems and emphasize the necessity of a unified theoretical approach.

In this direction, we expect that the inclusion of quantum effects may also, regularize or even eliminate the singularities present in those solutions of general relativity. That inclusion has motivated several semiclassical approaches as the Kazakov–Solodukhin metric, which describes a quantum deformation of the Schwarzschild solution where the backreaction of quantum fluctuations smooths out the central singularity, resulting in a regular and geodesically complete spacetime \cite{Kazakov:1993ha}.

Following those ideas, in \cite{Wu:2022lqr} the Kazakov-Solodukhin approach is extended to charged and asymptotically Anti de-Sitter (AdS) spacetimes by constructing a quantum corrected AdS black hole with electric charge . In such a case, a new parameter $a$, associated with spherically symmetric quantum fluctuations, introduces a minimal length scale and effectively weakens the electric charge at small radii. This modification alters both the causal structure and the thermodynamical properties of the black hole. 

In the present work we will be concerned with perturbations delivered by different fields in \textcolor{black} {AdS and dS black holes} and the stability question they pose. In most scenarios those perturbations evolve as \textcolor{black}{complex} frequencies, the quasinormal modes. 

The study of quasinormal modes (QNMs) provides a very useful framework for the investigation of the dynamical response of black holes under small perturbations. Each mode is characterized by a complex pair of quantities, whose real part governs the oscillation frequency and whose imaginary part determines the damping of the evolution  \cite{Berti:2009kk,Konoplya:2011qq}. 
When this imaginary part is negative, the spacetime is considered linearly stable, implying that perturbations decay exponentially in time. \textcolor{black}{A positive imaginary part}, on the other hand,  signal the onset of instabilities that could alter the geometry. Consequently, the computation of QNMs constitutes a tool for assessing the stability of black hole solutions, both in GR and in alternative theories as the quantum-corrected framework we consider in this work. It is worth to mention still that in the context of gravitational wave astronomy, QNMs play a crucial observational role. The ringdown phase following the merger of compact objects is described by a superposition of damped oscillations associated with these modes, allowing one to extract key parameters of the remnant black hole, such as its mass and angular momentum and to test the validity of the no-hair theorem \cite{Abbott:2016blz}. Current and future observations from the LIGO–VIRGO–KAGRA (LVK) collaboration are expected to confirm that the dominant fundamental mode agrees with the predictions of GR, while the precise identification of higher overtones remains challenging. Future detections with improved sensitivity will enable a more complete reconstruction of the QNM spectrum, opening a window to probe subtle deviations that may arise e. g. from Lorentz symmetry breaking or quantum corrections in the gravitational sector \cite{Casana:2017jkc}.

\textcolor{black}{
Quasinormal modes of black holes arising from higher-curvature and higher-derivative corrections to general relativity have been extensively investigated. It has been shown that higher-derivative corrections to the Schwarzschild geometry can lead to measurable deviations in the quasinormal spectrum, even when such corrections are small \cite{Blazquez-Salcedo:2016enn}. Related analyses were later performed for black holes in Einstein-dilaton-Gauss-Bonnet gravity, where the presence of higher-curvature terms modifies both the scattering properties of perturbations and the corresponding quasinormal frequencies \cite{Konoplya:2019}. Further investigations explored perturbations of rotating black holes in higher-derivative gravity frameworks, demonstrating that effective field-theory corrections may influence the ringdown signal and the associated stability properties \cite{Cano:2020}.
}

\textcolor{black}{
Additional developments in this direction have considered quasinormal spectra in string-corrected black hole geometries. In particular, the eikonal quasinormal modes and shadow properties of string-corrected higher-dimensional black holes were investigated in \cite{Moura:2021PLB}, while the asymptotic quasinormal spectrum of string-corrected black holes was obtained analytically using monodromy techniques in \cite{Moura:2021JHEP}. Complementary analyses of the asymptotic quasinormal frequencies in higher-derivative gravity theories were presented in \cite{Moura:2023}, and the sensitivity of higher overtones to such corrections was further explored in Einstein-Weyl gravity in \cite{Konoplya:2023}. Although the deformation considered in the present work originates from quantum corrections to the Schwarzschild metric \cite{battista1, battista2}, these studies share the common goal of understanding how departures from classical general relativity can manifest themselves in the quasinormal spectrum of black holes.
}

From a theoretical perspective, quasinormal spectra also encode information about how modifications to spacetime geometry influence stability and wave propagation. In quantum inspired metrics \cite{saghafi1} such as the Kazakov–Solodukhin black hole \cite{Kazakov:1993ha}, the presence of a minimal length scale alters the effective potential and shifts the QNM frequencies \cite{Konoplya:2019xmn,Bolokhov:2023ozp}, while in Bumblebee gravity \cite{Casana:2017jkc}, the Lorentz-violating parameter introduces distinctive signatures in both the oscillation and damping times. These changes can lengthen the ringdown phase or generate nearly resonant modes. Thus, the analysis of QNMs serves not only as a stability criteria but also as a potential observational probe for distinguishing among different modified gravity scenarios \cite{moyses24}.

In this work, we study the quantum corrected black hole with a cosmological constant, in both AdS and dS backgrounds. We compute its QNMs to examine the impact of the quantum correction parameter $a$ on the spectrum, emphasizing the modifications induced by the interplay between the quantum parameter, $a$ and the cosmological constant, $\Lambda$. 

The paper is organized as follows. Section (\ref{sec:back}) reviews the main ideas of the Kazakov-Solodukhin method \cite{Kazakov:1993ha} and introduces the quantum-corrected black hole solution with a cosmological constant. In Section (\ref{sec:pert}), we present the fundamental equations governing the dynamics of scalar and Dirac perturbations. Section (\ref{sec:meth}) describes the two numerical methods employed to compute the quasinormal spectrum. The results are reported in Section (\ref{sec:nume}), and Section (\ref{sec:disc}) contains the discussion and concluding remarks.

\section{Background}
\label{sec:back}

It is widely recognized that general relativity is  non-renormalisable and, therefore, cannot be consistently quantized within the framework of perturbative quantum field theory, see \cite{Shomer:2007vq}. In this section we turn to approaches that attempt to overcome partially this limitation.

In the special case in which the Einstein-Hilbert action can be reduced to a two-dimensional dilaton gravity considering spherical symmetry, a non-trivial feature is presented: the spacetime element includes a quantum correction.
In this regard, Kazakov and Solodukhin originally presented a modified version of the Schwarzschild black hole corrected by terms of quantum nature (see \cite{Kazakov:1993ha} for further details).
Since then, new solutions have been discovered in the field and in particular, as e. g. a charged geometry with non-vanishing cosmological constant, recently studied in \cite{Wu:2022lqr}, where quantum fluctuations in an Reissner-Nordström (RN) version of black hole were studied. Afterwards, an AdS spacetime background was included with the analysis of a quantum-corrected Reissner-Nordström-Anti-de Sitter (RNAdS) black hole.

It is generally expected that any quantum-correction should be very small. This motivates the ongoing search for subtle quantum signatures on classical backgrounds. Put differently, any departure from classical black hole solutions can only appear as a small modification of the available theories. 

Following this perspective and in light of the relevance of quantum features in the context of black hole physics, we investigate in this work the quasinormal modes of different geometries in which a quantum parameter is introduced and establish whether the background is stable against two types of perturbation. We will focus on more realistic physical scenarios in which the electric charge is set to zero, i.e. $Q = 0$, for a small value of the quantum parameter $a$. Although the charged case is a natural generalization, geometries with a positive or negative cosmological constant are more important, from an astrophysical or theoretical point of view (viz. AdS/CFT). For such reason in this work we investigate cosmological perturbed backgrounds.
\footnote{It should be noted that in our simplest case, $a/r << 1$, an artificial charge naturally arises from the underlying quantum background. We will present this result in the last equation of this section.}

To make our discussion self-contained, we will briefly review the most relevant equations required to construct the quantum-inspired black hole background. The departure idea is quite simple: based on the Einstein-Hilbert action with a cosmological constant in four dimensions, we link the GR solution to its equivalent in a two-dimensional spacetime using the expression \cite{Wu:2022lqr}
\begin{align}
    R &= R^{(2)} + \frac{2}{r^2} \nabla_\alpha r \nabla^\alpha r + \frac{2}{r^2} - \frac{2}{r^2} \nabla_\alpha \nabla^\alpha r^2,
\end{align}
in which the parameters have their usual meaning, i. e., $R$ is the Ricci scalar in four-dimensions, $R^{(2)}$ is the Ricci scalar in two-dimensions and $\Lambda$ is the cosmological constant.
Following the method developed in \cite{Kazakov:1993ha} and considering $r$ as a space-like coordinate, the Ansatz to be studied in the previous theory is that of a spherically symmetric line-element expressed as
\be
\label{metric}
\mathrm{d}s^2 &= -f(r) \, \mathrm{d}t^2 + f(r)^{-1} \, \mathrm{d}r^2 + r^2 \left( \mathrm{d}\theta^2 + \sin^2\theta \, \mathrm{d}\phi^2 \right),
\ee
that substituted in the motion equations yields the lapse function
\begin{align}
\label{me1}
f(r) &= -\frac{2M}{r} + \frac{1}{3}\Lambda r^2 + \frac{1}{r} \int U(r) \, \mathrm{d}r, 
\end{align}
in which $M$ is the black hole mass and $\Lambda$ is the cosmological constant.

\textcolor{black}{
It is worth emphasizing that the function $U(r)$ emerges inspired from the 
dimensional reduction of four-dimensional Einstein gravity under spherical symmetry, which leads to an effective two-dimensional dilaton gravity theory. 
In this approach, the radial function $r$ plays the role of a dilaton field, and quantum corrections associated with spherically symmetric fluctuations renormalize the dilaton potential, resulting in a generalized function $U(r)$. This procedure follows the original construction in \cite{Kazakov:1993ha}, where the four-dimensional Einstein-Hilbert action, when restricted to spherically symmetric configurations, reduces to an effective two-dimensional dilaton gravity 
model. In this framework, quantum effects modify the dilaton potential, while preserving the static and spherically symmetric character of the spacetime. The classical Schwarzschild-(A)dS solution is obtained in the limit $U(r)=1$, whereas deviations from this value gives effect of quantum corrections on the geometry.}

\textcolor{black}{
Finally, it is essential to highlight that, as Kazakov and Solodukh demonstrated in Ref.~\cite{Kazakov:1993ha}, quantum corrections profoundly change the form of the effective potential $ U(\phi(r)) \equiv U(r) $. This is precisely why the inclusion of such a correction term provides a consistent and satisfactory incorporation of quantum effects within the framework under consideration.
}

By applying corrections to the gravitational part, we use $U(r)$ as brought by \cite{Kazakov:1993ha},
\begin{align}
U(r) = \frac{r}{\sqrt{r^2 - a^2}}. 
\end{align}
\textcolor{black}{The parameter $a$ determines the minimal radius $r=r_{\text{min}}$ of the spacetime, so that the classical singularity at $r=0$ is replaced by $r=a$. Accordingly, the radial coordinate is restricted to $r>a$.}
\textcolor{black}{
To be more precise, and in accordance with the literature, the minimum length $a$ is related to the gravitational constant $\kappa$ (or, equivalently, to the Planck length $\ell_{\rm P}$) through the relation $a \equiv 4\sqrt{\kappa} = 4\ell_{\rm P}$.
}

The above derivation yields a simplified expression to the metric lapse function corresponding to its final form
\begin{align}
f(r) = \sqrt{1 - \frac{a^2}{r^2}} - \frac{2M}{r} - \frac{1}{3}\Lambda r^2. 
\label{lap1}
\end{align}
with an extra parameter $a$ when compared to usual \textcolor{black}{AdS and dS black holes}.

\textcolor{black}{Notice that the metric is defined only for $r \ge r_{\min}$, where $r_{\min}$ is the minimal radius determined by the effective potential $U(r)$. This reflects the fact that quantum corrections replace the classical singularity at $r=0$ by a minimal radius surface, beyond which the spacetime is not described by the effective geometry considered here.}

We still notice that the exact black hole solution (with $Q \neq 0$) was reported in \cite{Wu:2022lqr}. 
In this case, the critical points are obtained from a third-degree polynomial equation when $f(r_{critical}) = 0$. Considering positive roots ($r > 0$) there are at least two possible solutions: the cosmological horizon $r_{\Lambda}$ (for $\Lambda$ positive), which corresponds to the largest solution and the event horizon $r_h$, corresponding to the next one. The simplified algebraic equation is given explicitly by 
\begin{align}
    \Lambda ^2 r_c^6 
    + 
    12 \Lambda  M r_c^3
    - 
    9 r_c^2 +
    9 a^2 + 36 M^2
    &= 0.
\end{align}
The equation produces a non-trivial expression with six different roots for $f$, what can only be found numerically.

As we mentioned above, any quantum effect as the $a$ parameter we introduced in (\ref{lap1}) should be small. 

\textcolor{black}{Therefore, we employ the condition $a/r \ll 1$ as an approximation to solve the equation $f(r)=0$.} In this approximation, the lapse function can be expanded to a simple form,
\begin{align}
    f(r) \approx Y(r) \equiv 1 - \frac{2M}{r} - \frac{1}{3}\Lambda r^2 - \frac{(a^2/2)}{r^2} + \mathcal{O}(a^3),
\end{align}
where, by defining $q^2 \equiv a^2/2$ and interpreting the negative sign as corresponding to a phantom-like charged black hole, the original solution can, under this approximation, be regarded as a phantom \textcolor{black}{{\it AdS and dS Reissner-Nordstr\"om}} black hole \cite{Jardim:2012se,Liu:2023lwv}.
The approximated expression $Y(r_c) = 0$ gives us a more manageable expression,
\begin{align}
   \Lambda  r_c^4-3 r_c^2 +6 M r_c+\frac{3 a^2}{2}= 0,
\end{align}
where four critical points are present, being one of them before the event horizon, near the physical singularity. For small values of $a$, the roots obtained from the original lapse function $f(r)$ and those obtained from the approximated lapse function $Y(r)$ are essentially the same.

As mentioned in the preceding paragraphs, this manuscript investigates the influence of a quantum parameter in the dynamical stability of first-order perturbations. To this end, we develop the main ideas in the next section.

\section{Perturbations}
\label{sec:pert}
Our aim in this work is to investigate different field perturbations in black hole geometries that arise from a quantum-corrected version of general relativity. Here we provide an introductory revision of two distinct spin equations, the scalar one and the Dirac fields with their particular behavior, obtaining for each case, the motion equation as a one dimensional wave with a typical potential barrier. We start with the Klein-Gordon field after which we review the Dirac field, both cases without mass.

\subsection{Scalar perturbations}\label{sp1}

In order to examine the dynamics of a test scalar field, denoted by $\Phi$, propagating in a four-dimensional spacetime within the above gravitational background, we first consider the matter action, $S[g_{\mu \nu}, \Phi]$, 
\begin{align}
S[g_{\mu \nu} ,\Phi] \equiv \frac{1}{2} \int \mathrm{d}^4 x \sqrt{-g}
\Bigl[
\partial^{\mu} \Phi \partial_{\mu} \Phi 
\Bigl]\, ,
\end{align}
that provides the motion equation for the scalar field once consider to be an invariant \cite{Crispino:2013pya,Kanti:2014dxa,Pappas:2016ovo,Panotopoulos:2019gtn,Avalos:2023ywb,Gonzalez:2022ote,Rincon:2020cos},
\begin{equation}
\frac{1}{\sqrt{-g}}\partial_{\mu}\left(\sqrt{-g}g^{\mu\nu}\partial_{\nu}\Phi\right) = 0.
\end{equation}
We can exploit the symmetries of the metric by introducing a spherical-like Ansatz for the field decomposition considering an angular space spanned by spherical harmonics as eigenfunctions, what is done through the sum over all possible eigenvalues, 
\begin{equation}
\Phi(t, r, \theta, \phi) 
=\sum_{\ell ,m}e^{-i\omega t}\frac{\psi(r)}{r}Y_{\ell m}(\theta, \phi).\label{fdc}
\end{equation}
%
After applying the above Ansatz the differential equation can be cast into the form
\begin{align}
\begin{split}
& \frac{\omega^{2}r^{2}}{f(r)} + \frac{r}{\psi(r)}\frac{\mathrm{d}}{\mathrm{d}r}\left[r^{2}f(r)\frac{\mathrm{d}}{\mathrm{d}r}\left(\frac{\psi(r)}{r}\right)\right] +
\\
&\frac{1}{Y(\Omega)}\left[\frac{1}{\sin\theta}\frac{\partial}{\partial\theta}\left(\sin\theta\frac{\partial Y(\Omega)}{\partial\theta}\right)\right] +
\frac{1}{\sin^{2}\theta}\frac{1}{Y(\Omega)}\frac{\partial^{2}Y(\Omega)}{\partial\phi^{2}} 
= 0.
\label{KG}
\end{split}
\end{align}

At this point, we can recognize the angular part and replace it to the correspondent eigenvalues, i.e., 
\begin{align}
    \begin{split}
&\frac{1}{\sin\theta}\frac{\partial}{\partial\theta}\left(\sin\theta\frac{\partial Y(\Omega)}{\partial\theta}\right) + \frac{1}{\sin^{2}\theta}\frac{\partial^{2}Y(\Omega)}{\partial\phi^{2}} = 
-\ell(\ell + 1)Y(\Omega),
\label{kg2}
\end{split}
\end{align}
where $\ell(\ell + 1)$ is the eigenvalue, and $\ell$ is the angular degree (i. e. the field angular momentum). By combining (\ref{KG}) and (\ref{kg2}), we obtain a second-order differential equation for the radial coordinate 
To do that we will introduce the so-called "tortoise coordinate" $r_{*}$ usually defined as
\be
\label{tcd1}
    dr_{*}  \equiv \frac{\mathrm{d}r}{f(r)}\,, \hspace{2.5cm} f(r) \partial_r = \partial_{r_*}
\ee
and then rewrite the resulting differential equation in its Schr{\"o}dinger-like form
\begin{equation} \label{SLE}
\frac{\mathrm{d}^{2}\psi(r_*)}{\mathrm{d}r_{*}^{2}} + \left[\omega^{2} - V_{S}(r)\right]\psi(r_*) = 0,
\end{equation}
where $V_{S}(r)$ is the corresponding effective potential given by
\begin{equation}
V_{S}(r) = f(r)
\Bigg[ 
\frac{\ell(\ell + 1)}{r^{2}} + \frac{f'(r)}{r}
\Bigg]\label{poten}
\end{equation}
and the prime denotes the derivative of the radial variable. 

\begin{figure*}[ht!]
     \centering
     \begin{subfigure}[b]{0.34\textwidth}
         \centering
         \includegraphics[width=\textwidth]{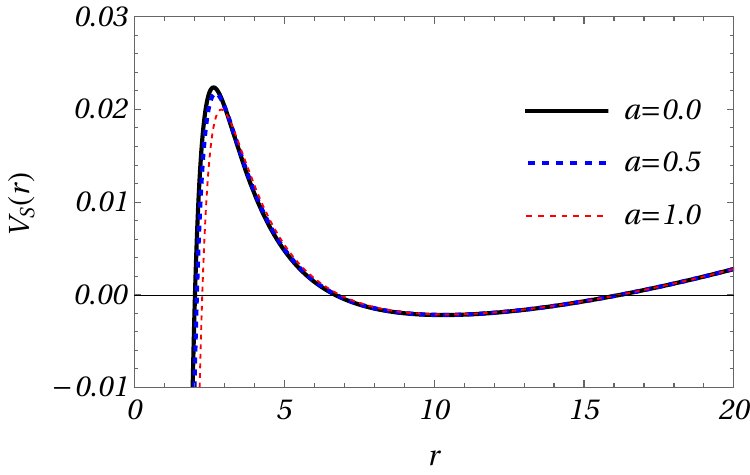}
         \label{Fig1SLeft}
     \end{subfigure}
     \hfill
     \begin{subfigure}[b]{0.32\textwidth}
         \centering
         \includegraphics[width=\textwidth]{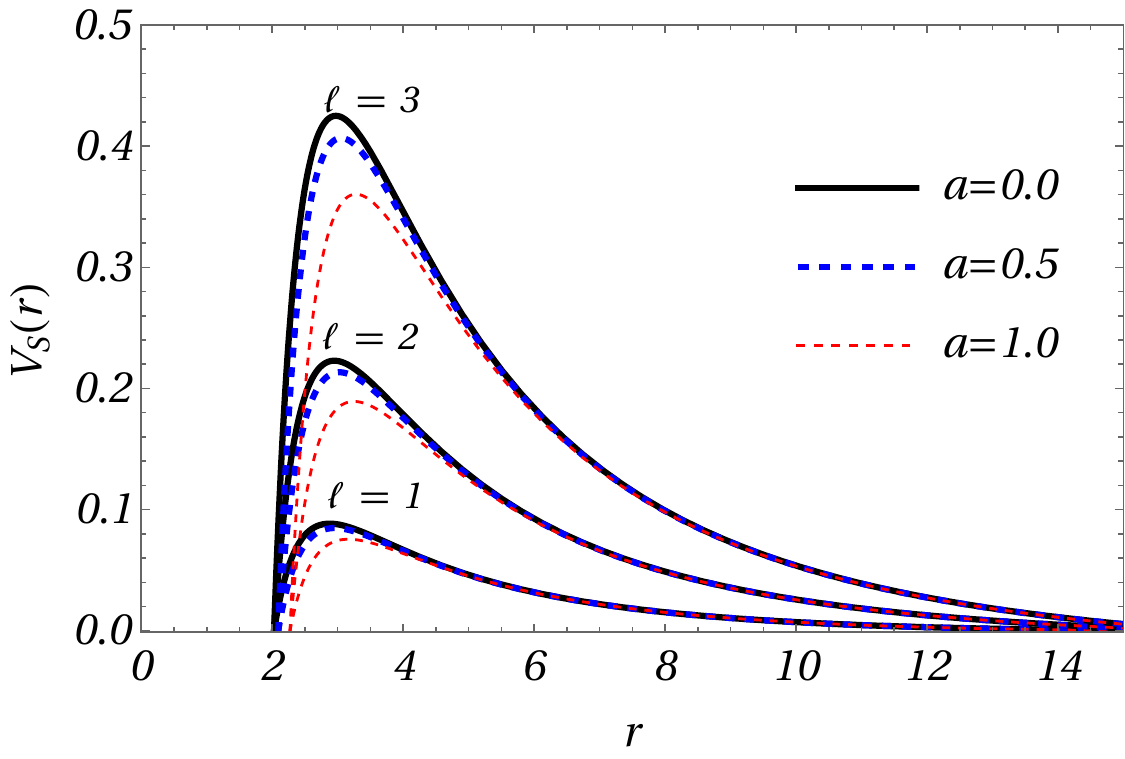}
         \label{Fig1SRight}
     \end{subfigure}
        \begin{subfigure}[b]{0.32\textwidth}
         \centering
         \includegraphics[width=\textwidth]{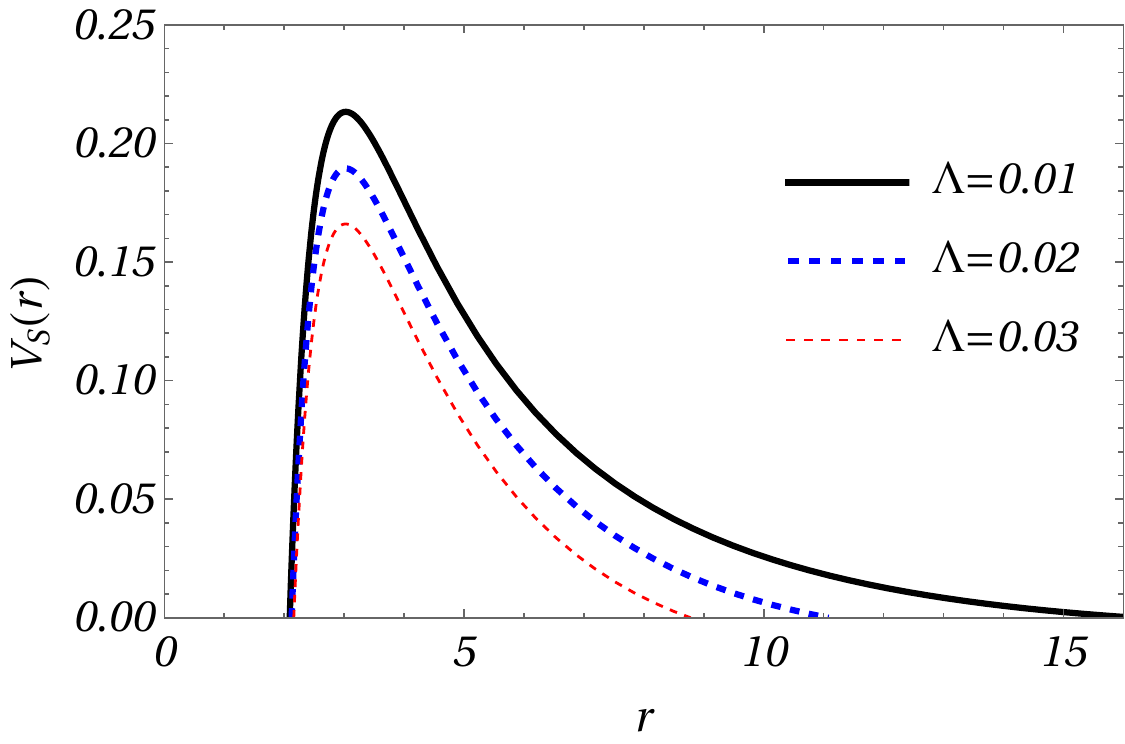}
         \label{Fig2SLeft}
     \end{subfigure}
      \caption{\footnotesize Effective potential $V_{S}$ for scalar perturbations in the \textbf{de Sitter} case, with a fixed mass $M=1$ and different values of $\{a, \ell, \Lambda\}$. \textit{Left panel:} Varying the quantum-correction parameter $a$ with \textcolor{black}{$\ell = 0$} and $\Lambda = 0.01$. \textit{Middle panel:} Varying the multipole number $\ell$ for three different values of $a$ with $\Lambda = 0.01$. \textit{Right panel:} Varying the cosmological constant $\Lambda$ with $\ell = 2$ and $a = 0.5$.}

        \label{fig:1}
\end{figure*}

In Fig.~(\ref{fig:1}) we display the behavior of the effective potential $V_S(r)$ for scalar perturbations in the de Sitter case. For the lowest multipole number, $\ell = 0$, one can notice the presence of a region where the potential becomes negative, which may indicate the possibility of unstable modes depending on the boundary conditions. In all cases, the potential exhibits a single peak located close to the event horizon and smoothly decays to zero as $r$ approaches the cosmological horizon. Moreover, the inclusion of the quantum correction parameter $a$ tends to make the potential barrier less pronounced, effectively lowering its maximum height. This feature reflects the role of quantum effects in softening the potential structure near the black hole, which may have direct implications for the quasinormal spectrum and the stability analysis.

\begin{figure*}[ht!]
     \centering
     \begin{subfigure}[b]{0.33\textwidth}
         \centering
         \includegraphics[width=\textwidth]{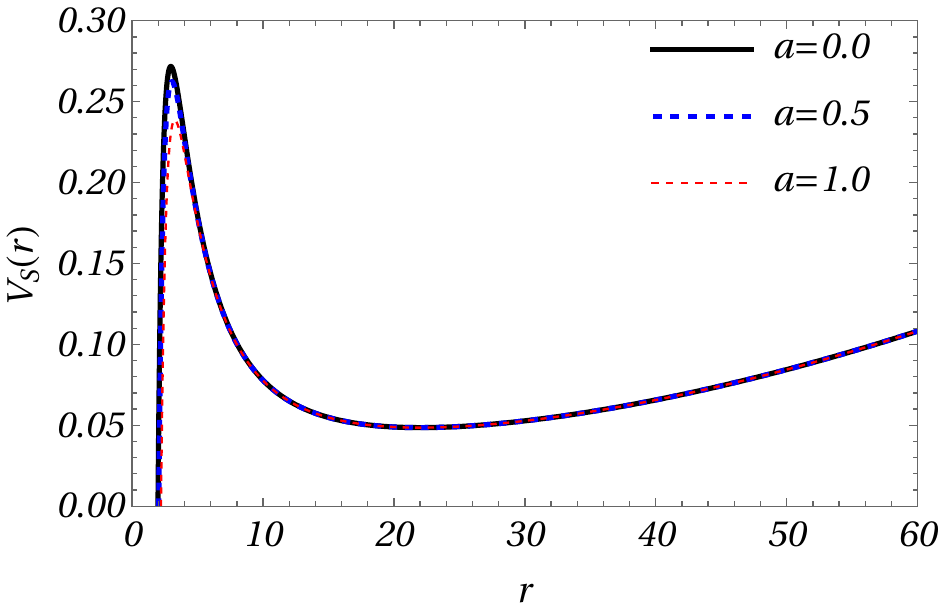}
         \label{Fig1SLeft}
     \end{subfigure}
     \hfill
     \begin{subfigure}[b]{0.32\textwidth}
         \centering
         \includegraphics[width=\textwidth]{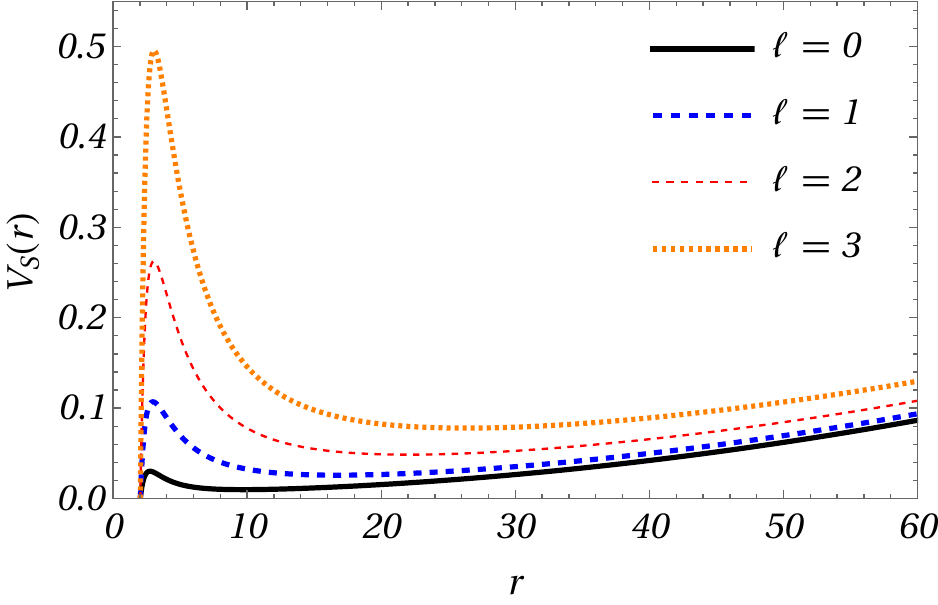}
         \label{Fig1SRight}
     \end{subfigure}
        \begin{subfigure}[b]{0.32\textwidth}
         \centering
         \includegraphics[width=\textwidth]{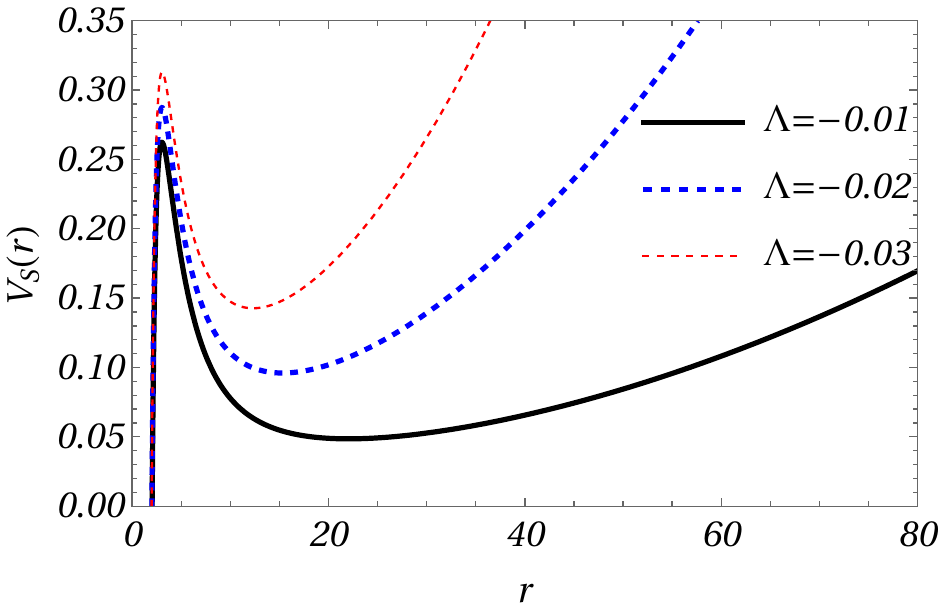}
         \label{Fig2SLeft}
     \end{subfigure}
      \caption{\footnotesize Effective potential $V_{S}$ for scalar perturbations in the \textbf{Anti-de Sitter} case, with a fixed mass $M=1$ and different values of $\{a, \ell, \Lambda\}$. \textit{Left panel:} Varying the quantum-correction parameter $a$ with $\ell = 2$ and $\Lambda = -0.01$. \textit{Middle panel:} Varying the multipole number $\ell$ \textcolor{black}{with $a=0.5$}, $\Lambda = -0.01$. \textit{Right panel:} Varying the cosmological constant $\Lambda$ with $\ell = 2$ and $a = 0.5$.}

        \label{fig:2}
\end{figure*}

Figure~(\ref{fig:2}) shows the effective potential $V_S(r)$ for scalar perturbations in the Anti–de Sitter (AdS) background. In this case, the potential also develops a peak close to the event horizon, but unlike the de Sitter geometry, it grows without bound as $r$ increases, a typical feature of the confining AdS spacetime. The inclusion of the quantum correction parameter $a$ reduces the height and sharpness of the barrier, making the potential smoother near the horizon.

%

After the above simplifications, the determination of the quasinormal frequencies relies on imposing appropriate boundary conditions, which generally depends on the wave behavior in the asymptotic regions of the spacetime. Typically, for dS and plane asymptotic regions, we have
\be
  \Phi \rightarrow \: &\exp(-i \omega r_*), \; \; \; \; \; \;  r_* \rightarrow - \infty ,
\label{pbc1}
   \\
   \Phi \rightarrow \: &\exp(+i \omega r_*), \; \; \; \; \; \; r_* \rightarrow + \infty .
\label{pbc}
\ee

That is not the case in AdS-like spacetimes, since the potential is usually divergent for $r\rightarrow \infty$ and $r_*$ does not hit $+\infty$. \textcolor{black}{Considering that (\ref{tcd1}) is not analytical, we can still investigate the AdS boundary by expanding $f$. To leading order we have 
\be
\lim_{r \rightarrow \infty } r_* \rightarrow \mathrm{constant} + \frac{3}{\Lambda r} + \mathscr{O}(r^{-2})\simeq \mathrm{constant}.
\ee
in which the constant can be set to zero as a coordinate gauge fixing. Then, in} AdS boundaries we have
\be
 & \Phi & \rightarrow \: \exp(-i \omega r_*), \; \; \; \; \; \;   r_*  \rightarrow - \infty ,
\label{pbc2}   \\
  & \Phi & \rightarrow \: 0, \; \; \; \; \; \; \; \; \; \; \; \; \; \; \; \; \; \; \; \; \; \;  r_*  \rightarrow 0 .
\label{pbc3}
\ee

%

It is worth recalling that, by assuming a time dependence of the form $\Phi \sim e^{-i \omega t}$, as expressed in Eq.~(\ref{fdc}), a negative imaginary part of the frequency $\omega$ corresponds to a decaying (stable) mode, whereas a positive imaginary part indicates a growing (unstable) mode.

The quasinormal frequencies or the eventual instabilities are completely determined by the effective potential (\ref{poten}) of the field. When this function is at least partly negative (as e. g. in de Sitter spacetimes) \cite{Horowitz_2000} the field may evolve with increasing amplitude in time, denoting an unstable evolution that eventually changes the geometry.

\subsection{Dirac perturbations}

We now present the basic framework for obtaining the motion equation of the neutral Dirac particles and calculate the quasinormal modes of such field. To begin, we consider a spherically symmetric background as disposed in (\ref{metric})  (for further details see e. g.  \cite{Cho:2007zi}).  We start by performing a conformal transformation in (\ref{metric})  as (see \cite{Das:1996we,Gibbons:1993hg} and references therein)
\begin{eqnarray}
g_{\mu\nu} & \rightarrow & \overline{g}_{\mu\nu}=\Omega^{2}g_{\mu\nu} , \\
\psi & \rightarrow & \overline{\psi}=\Omega^{-3/2}\psi , \\
\gamma^{\mu}\nabla_{\mu}\psi & \rightarrow & \Omega^{5/2} \overline{\gamma}^{\mu}\overline{\nabla}_{\mu}\overline{\psi} ,
\end{eqnarray}
with $\overline{\psi}=r^{3/2}\psi$ and the conformal factor, $\Omega=1/r$. The line element  is then casted in the following form
\be
\label{metric2}
\mathrm{d}\overline{s}^{2} = -\frac{1}{r^{2}}f(r)\mathrm{d}t^{2} + \frac{1}{r^{2}}f(r)^{-1}\mathrm{d}r^{2} + \mathrm{d}\Omega^{2}. 
\ee

Now, let us take advantage of the fact that we can split the $(t,r)$ part from the $2$-sphere what allows us to decouple the equation to study the radio-temporal part. Thus, writing down the Dirac perturbation in the form (massless particles)
\begin{eqnarray}\nonumber
\label{diraceq}
\overline{\gamma}^{\mu}\overline{\nabla}_{\mu}\overline{\psi}  &=& 
\bigg[ 
\left(
\overline{\gamma}^{t}\overline{\nabla}_{t} +
\overline{\gamma}^{r}\overline{\nabla}_{r} \right) \otimes 1
+ 
\overline{\gamma}^{5} \otimes
\left( \overline{\gamma}^{a}
\overline{\nabla}_{a}\right)_{S_{d-2}} 
\bigg] 
\overline{\psi} \\
&=& 0 , 
\end{eqnarray}
in which, as usual $(\overline{\gamma}^{5})^{2}=1$, and the field Ansatz,
\begin{equation}
\overline{\psi} = \sum_{\ell} \left( \phi_{\ell}^{(+)} (r,t) \chi_{\ell}^{(+)} (\theta , \phi ) + \phi_{\ell}^{(-)} (r,t) \chi_{\ell}^{(-)} (\theta , \phi ) \right) .
\end{equation}
we draw upon the fact that the angular part has the fixed eigenvalues, $\pm i( \ell +1)$ (hereby defined as $\pm i \xi$) and consider $\chi_{\ell}^{(\pm)}$ as the eigenspinors of the $2$-sphere (see \cite{Camporesi:1995fb} for further details),
\be
\left( \gamma^{a}\nabla_{a} \right)_{S_2}\chi_{\ell}^{(\pm)} = \pm i \left( \ell + 1\right) \chi_{\ell}^{(\pm)}.
\ee

Here $\ell = 0, 1, 2, \dots$ and the equation (\ref{diraceq}) is rewritten (omitting the bars for simplicity) as
\begin{equation}
\label{eqn:2Ddirac}
\left \{ 
\gamma^{t} \nabla_{t} + \gamma^{r} \nabla_{r} + \gamma^{5} \left[ \pm i \left( \ell + 1 \right) \right] \right \} 
\phi_{\ell}^{(\pm)} = 0 ,
\end{equation}
which is precisely the pair of $2$-dimensional Dirac equations in coordinates $r$ and $t$.
In order to solve the differential equations we make an explicit choice of the Dirac matrices, namely
\begin{align}
\gamma^{t} &= \frac{r}{\sqrt{f(r)}}(-i\sigma^{3}), 
\\
\gamma^{r} &=  \sqrt{f(r)} 
\ r\sigma^{2} ,
\end{align}
in which $\sigma^{i}$ are the so-called Pauli matrices, defined as
\begin{equation}
\sigma^{1}=\left(
\begin{array}{cc}
0 & 1 \\ 1 & 0
\end{array}
\right)\ \ \ ,\ \ \ \sigma^{2}=\left(
\begin{array}{cc}
0 & -i \\ i & 0
\end{array}
\right)\ \ \ ,\ \ \ \sigma^{3}=\left(
\begin{array}{cc}
1 & 0 \\ 0 & -1
\end{array}
\right) .
\end{equation}

Also, $\gamma^{5}$ is written in terms of the Pauli matrices as follows
\begin{equation}
\gamma^{5} = (-i\sigma^{3})(\sigma^{2}) = - \sigma^{1} .
\end{equation}

The spin connections are then found to be:
\begin{align}
\Gamma_{t} &= \sigma^{1}  
\frac{r^2}{4} \frac{\mathrm{d}}{\mathrm{d}r} \left( \frac{f(r)}{r^{2}} \right) , \\
\Gamma_{r} &= 0 .
\end{align}

At this point, it should be mentioned that the treatment for the $+$ signal solution is completely equivalent to the $-$ signal case. For such reason we focus on the positive case, casted as

\begin{align}
\begin{split}
 \left\{
 \frac{r}{\sqrt{f(r)}}(-i\sigma^{3}) \left[ \frac{\partial}{\partial t} + \sigma^{1}\frac{ r^{2}}{4} \frac{\mathrm{d}}{\mathrm{d}r} \left( \frac{f(r)}{r^{2}} \right)
\right] \right.
+\\ 
\left.
 \sqrt{f(r)} r \sigma^{2} \frac{\partial }{\partial r} + (-\sigma^{1})(i) \left( \xi \right) \right\} \phi_{\ell}^{(+)} = 0.  
\end{split}
\end{align}

Writing the spatial and temporal part separately, a pair of first order partial differential equation emerges for $\phi_{\ell}^{(+)}$, i.e., 

\begin{align}
\begin{split}
\sigma^{2} 
\left( \sqrt{f(r)} \ r\right) 
\left[ \frac{\partial}{\partial r} + \frac{r}{2\sqrt{f(r)}}
\frac{\mathrm{d}}{\mathrm{d}r} \left( \frac{\sqrt{f(r)}}{r} \right) \right]
\phi_{\ell}^{(+)}\\ - i \sigma^{1} \xi 
\phi_{\ell}^{(+)} = i \sigma^{3} \left( \frac{r}{\sqrt{f(r)}} \right)
\frac{\partial \phi_{\ell}^{(+)}}{\partial t} . &
\end{split}
\end{align}
To integrate the above relation we must set the spinor $\phi_\ell$ in terms of two different components,
\begin{equation}
\phi_{\ell}^{(+)} = \left( \frac{\sqrt{f(r)}}{r} \right)^{-1/2} e^{-i \omega t} \left(
\begin{array}{c}
i\mathcal{G} (r) \\ \mathcal{F} (r)
\end{array}
\right) ,
\end{equation}
%
%
reducing the Dirac equation to
\begin{equation}
\begin{split}
\sigma^{2} \left(  \sqrt{f(r)} \ r \right) \left(
\begin{array}{c}
i\frac{\mathrm{d} \mathcal{G} (r)}{\mathrm{d}r} \\ 
\ \frac{\mathrm{d} \mathcal{F} (r)}{\mathrm{d}r}
\end{array}
\right) -i \sigma^{1} \xi \left(
\begin{array}{c}
i \mathcal{G} (r) 
\\ 
\mathcal{F} (r)
\end{array}
\right) = 
\\ 
\sigma^{3} \omega \left( \frac{r}{\sqrt{f(r)}} \right) \left(
\begin{array}{c}
i \mathcal{G} (r) 
\\ 
\mathcal{F} (r)
\end{array}
\right) .
\end{split}
\end{equation}

By taking the components separately, we finally write a set of coupled first-order differential equations in terms of the variables $\mathcal{G}  \equiv \mathcal{G} (r)$ and $\mathcal{F} \equiv \mathcal{F} (r)$ as follows
\begin{eqnarray}
f(r) \frac{\mathrm{d} \mathcal{G} (r)}{\mathrm{d}r} 
- 
\Bigg [ \frac{\sqrt{f(r)}}{r} \xi \Bigg ] \mathcal{G} (r) & = & + \omega \mathcal{F} (r)   \label{Eq1}
\\ 
f(r) \frac{\mathrm{d} \mathcal{F} (r)}{\mathrm{d}r} 
+ 
\Bigg [ \frac{\sqrt{f(r)}}{r} \xi \Bigg ] \mathcal{F} (r) & = & - \omega \mathcal{G} (r) . \label{Eq2}
\end{eqnarray}

Now, introducing the special potential $\mathcal{W} \equiv \xi \sqrt{f(r)}/r$ in terms of which, usually, the Dirac potentials are defined (as established bellow) and the tortoise coordinate defined in (\ref{tcd1}) we rewrite the pair of equations (\ref{Eq1}-\ref{Eq2}) to
\begin{align}
    \Bigg[
    \frac{\mathrm{d}}{\mathrm{d}r_{*}} - \mathcal{W} 
    \Bigg] \mathcal{G} &= + \omega\mathcal{F}
    \\
    \Bigg[
    \frac{\mathrm{d}}{\mathrm{d}r_{*}} + \mathcal{W} 
    \Bigg] \mathcal{F} &= - \omega \mathcal{G}.
\end{align}

This set of first-order differential equations for $\mathcal{G}$ and $\mathcal{F}$ can be easily decoupled to obtain two Schrodinger-like differential equations, with two concrete effective potentials, i.e.,
\begin{align}
\frac{\mathrm{d}^2\mathcal{F}}{\mathrm{d}{r_{*}}^2} + [\omega^2 - V_{-}] \mathcal{F} & =  0 , \label{SL1}
\\
\frac{\mathrm{d}^2 \mathcal{G}}{\mathrm{d}{r_{*}}^2} + [\omega^2 - V_{+}] \mathcal{G} & =  0 , \label{SL2}
\end{align}
where the potentials are given by 
\begin{equation}
\label{diracpot}
V_{D\pm} = \mathcal{W}^2 \pm \frac{\mathrm{d} \mathcal{W} }{\mathrm{d}r_{*}} = \mathcal{W}^2 \pm f(r) \frac{\mathrm{d} \mathcal{W} }{\mathrm{d}r}.
\end{equation}

\textcolor{black}{
Finally, it is worth emphasizing that, in the language of supersymmetry, the corresponding potentials $  V_{D+}  $ and $  V_{D-}  $ are expected to be superpartners. This follows from the fact that they can be derived from a common superpotential, which in turn implies that they yield identical spectra (see, e.g., Ref.~\cite{Cooper:1994eh} for further details).
}
We may see however spacetimes where such is not the case, justifying the break of the isospectrality property.

The explicit form of both effective potentials for the spacetimes we are considering are
\begin{eqnarray}\nonumber
\label{potex}
V_{D\pm} = \frac{(\ell+1)^2f(r)}{r^2} \pm (\ell+1)\sqrt{f(r)}\left( \frac{\partial_r f(r)}{2r} - \frac{f(r)}{r^2} \right).\\
\end{eqnarray}

\begin{figure*}[ht!]
     \centering
     \begin{subfigure}[b]{0.45\textwidth}
         \centering
         \includegraphics[width=\textwidth]{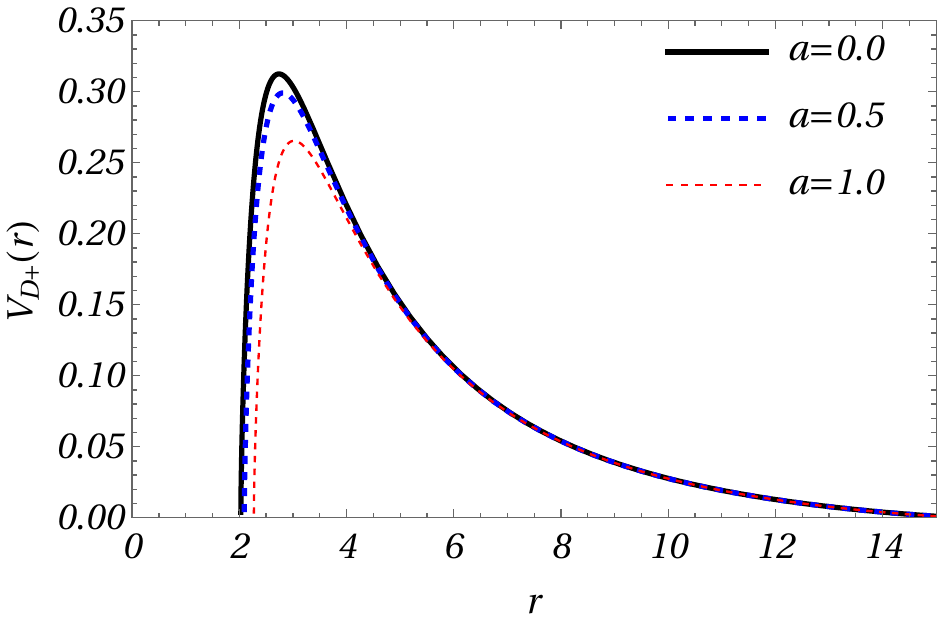}
         \label{Fig1SLeft}
     \end{subfigure}
     \begin{subfigure}[b]{0.45\textwidth}
         \centering
         \includegraphics[width=\textwidth]{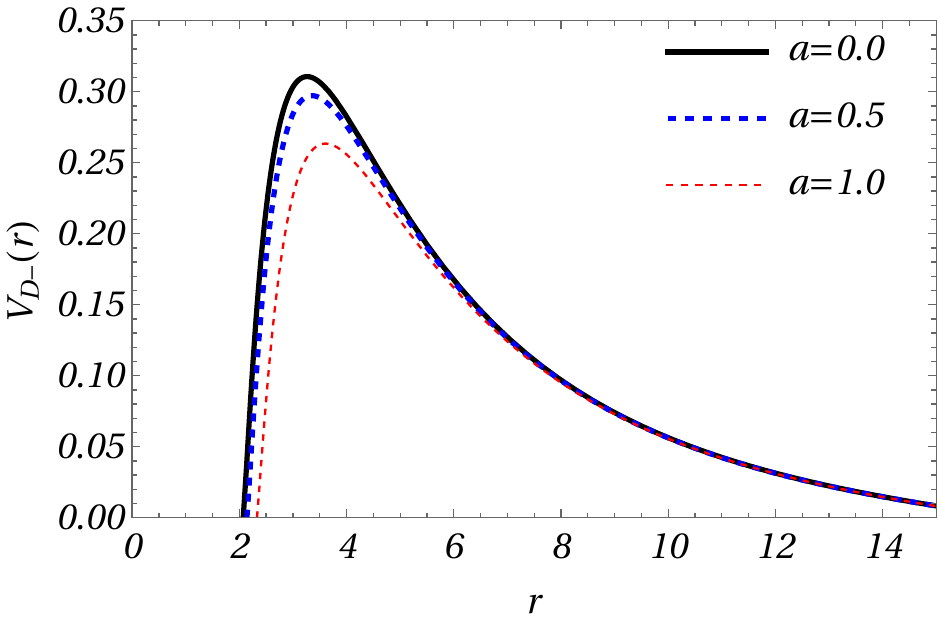}
         \label{Fig1SRight}
     \end{subfigure}
      \caption{\footnotesize Effective potential $V_{D\pm}$ for spinorial perturbations with different values of quantum-correction parameter $a$ in the \textbf{de Sitter} case, with a fixed mass $M=1$, $\ell = 2$ and $\Lambda = 0.01$.}

        \label{fig:3}
\end{figure*}

Figure (\ref{fig:3}) illustrates the effective potentials $V_{D+}(r)$ associated with the massless Dirac perturbations in the de Sitter background. Both $V_{D+}$ and $V_{D-}$ exhibit a single peak near the event horizon and gradually decay to zero as $r$ approaches the cosmological horizon. For $\ell = 2$, their overall profiles closely resemble those obtained in the scalar case, with $V_{D+}(r)$ and $V_{D-}(r)$ behaving as nearly supersymmetric partners. The quantum correction parameter $a$ reduces the height and steepness of the barrier. 

In Figure~4, we bring the Anti–de Sitter (AdS) spacetimes submitted to the Dirac field. The potentials display a pronounced maximum close to the event horizon and rise steadily for large $r$, reflecting the confining character of the AdS spacetime. Increasing the quantum correction parameter $a$ also lowers the barrier, slightly modifying the asymptotic behavior. Such alterations suggest that quantum effects also influence the spinorial sector, potentially affecting the isospectrality between $V_{D+}(r)$ and $V_{D-}(r)$ and the corresponding quasinormal spectra, as we may further see.

\textcolor{black}{The boundary conditions for the propagating Dirac field in the dS geometries are the same of the scalar case, (\ref{pbc1}-\ref{pbc}) since $V(r_h) = 0$ enabling Eq. (\ref{SL1}-\ref{SL2}) to admit the usual plane wave in asymptotic regions.}

\textcolor{black}{In the case of AdS geometries the analysis is more subtle. In the near-horizon vicinity, the same boundary condition as that of the scalar scattering can be applied since both equations are identical in such region (in both cases $V(r_h) =0$). The field equation, however, has a different boundary term in the AdS spacetime considering the Dirac potential (\ref{potex}), being expressed as
%
\be
\label{rch1}
\Psi \Big|_{r\rightarrow \infty } \rightarrow e^{-i r_* \sqrt{\omega^2 -V_\infty } }.
\ee}

\textcolor{black}{Since $r_* (r\rightarrow \infty ) = 0$ the above term reduces to a constant value that we take to be zero to be consistent with the field equation (\ref{SL1}-\ref{SL2}) in such region. Consequently, both conditions are the same for the Dirac and Scalar fields also in AdS spacetimes, expressed in (\ref{pbc2}-\ref{pbc3}).}


\begin{figure*}[ht!]
     \centering
     \begin{subfigure}[b]{0.45\textwidth}
         \centering
         \includegraphics[width=\textwidth]{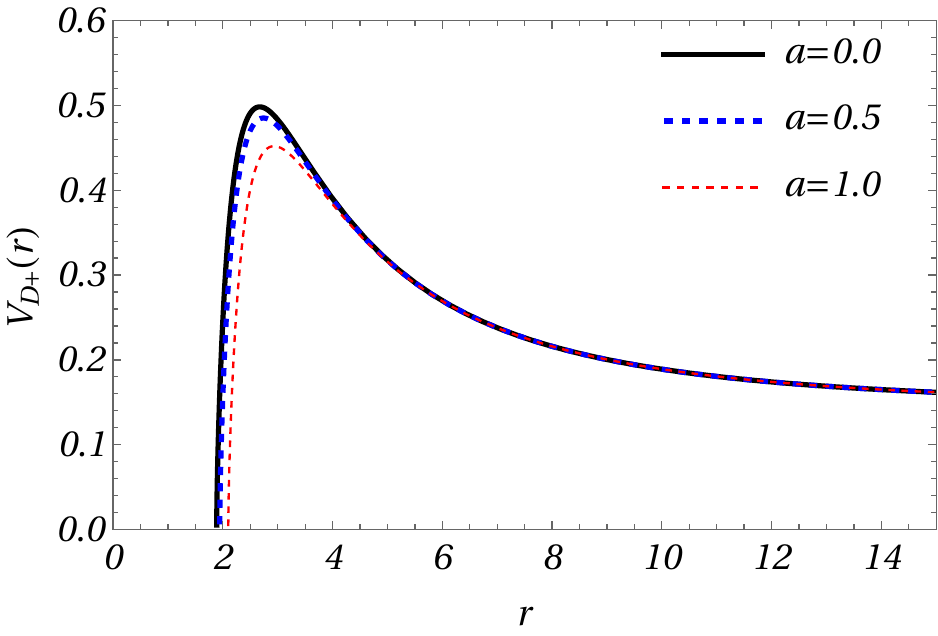}
         \label{Fig1SLeft}
     \end{subfigure}
     \begin{subfigure}[b]{0.45\textwidth}
         \centering
         \includegraphics[width=\textwidth]{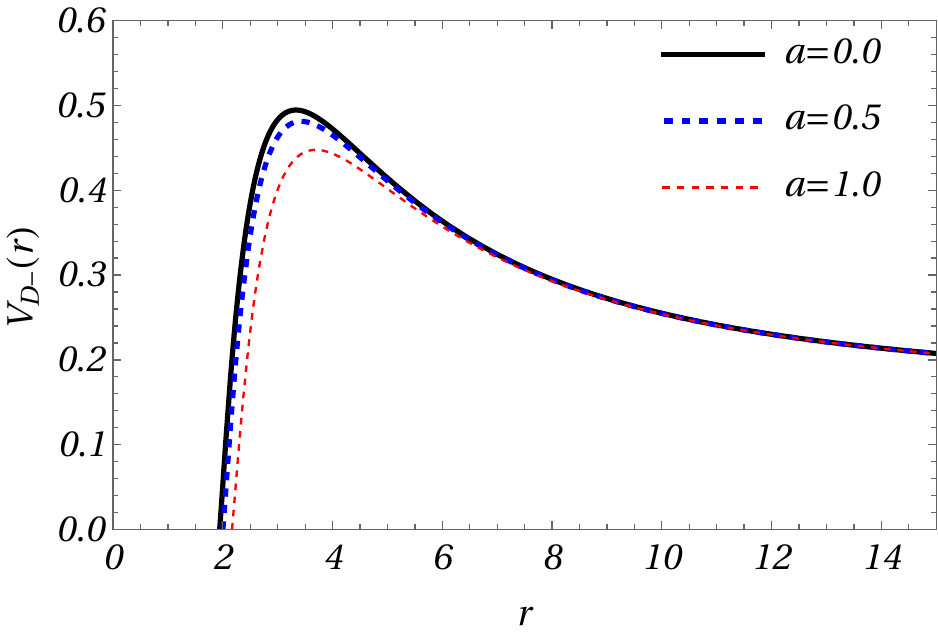}
         \label{Fig1SRight}
     \end{subfigure}
      \caption{\footnotesize Effective potential $V_{D\pm}$ for spinorial perturbations with different values of quantum-correction parameter $a$ in the \textbf{Anti-de Sitter} case, with a fixed mass $M=1$, $\ell = 2$ and $\Lambda = 0.01$.}

        \label{fig:4}

\end{figure*}

\section{Numerics}
\label{sec:meth}
For the quantum black holes we analyze in this manuscript, we employ two different methods for the study of field perturbations: the well-known WKB approach \cite{Schutz:1985km,Konoplya:2003ii,Konoplya:2003dd, Matyjasek:2017psv,Konoplya:2019hlu,Konoplya:2003dd,Konoplya:2002ky} and the characteristic integration in null coordinates as developed in \cite{Gundlach_1994, Konoplya:2011qq,Fontana_2021}. Since those methods were already extensively described in the specific literature of quasinormal modes, in this section we write a short review of their main features focusing on the novelty we employed.

\subsection{WKB semi-analytical approach}

The WKB approximation is an effective semi-analytical technique for determining the QNMs spectrum in the situations where the effective potential possesses a single, well-defined maximum in the vicinity of the black hole event horizon. Motivated by the formal analogy between the perturbation equations of black hole spacetimes and the one-dimensional Schrödinger equation with a potential barrier, the computation of QNMs frequencies can be recast as a scattering problem in quantum mechanics, where the boundary conditions correspond to purely ingoing waves at the event horizon and purely outgoing waves at spatial infinity (or at the cosmological horizon in asymptotically de Sitter spacetimes). 

In general, the perturbation equation can be put into a Schrödinger-like form  
\begin{equation}\label{wkb-schro1}
    \frac{d^{2} \Psi(x)}{dx^{2}} + Q(x)\,\Psi(x) = 0,
\end{equation}
where the coordinate $x$ is related to the tortoise coordinate $r_{*}$, which maps the spatial infinity and the event horizon to $+\infty$ and $-\infty$, respectively. The function $Q(x)$ is defined as
\begin{equation}
    Q(x) = \omega^{2} - V(x),
\end{equation}
with $\omega$ denoting the quasinormal frequency and $V(x)$ representing the effective potential barrier associated with the perturbation under consideration.

In the near-horizon region, the function $Q(x)$ is expanded in a Taylor series around $x = x_{0}$, where $x_{0}$ corresponds to the position of the maximum of the effective potential $V(x)$. At spatial infinity, the effective potential approaches either zero or a constant value, corresponding to asymptotically flat and asymptotically de Sitter spacetimes, respectively. Solving the perturbation equation~(\ref{wkb-schro1}) with the boundary condition of purely outgoing waves at $x \to +\infty$ and matching it to the near-horizon solution at $x \to -\infty$ with purely ingoing wave boundary conditions yields a quantization condition for the QNMs frequencies, $\omega = \omega_{n\ell}$, where $n$ is the overtone number and $\ell$ is the angular momentum eigenvalue, satisfying $\ell \geq n$ for convergent and numerical stable results.

\textcolor{black}{Following \cite{Schutz:1985km}, the QNMs boundary conditions are satisfied only when the solution selects the pole structure of the Gamma function, which occurs for integer values of its argument. Consequently, $n$ is restricted to be a non-negative integer. So, we have}

\begin{equation}
    \Gamma(-n) = \infty,
\end{equation}
where $\Gamma$ denotes the usual Gamma function, yielding
\begin{equation}\label{wkb-formula}
   n + \frac{1}{2} = i \frac{\omega^2 - V_0}{\sqrt{-2V_0''}} - \sum_{j=2}^{6} \Pi_j.
\end{equation}
The $V_0$ is the maximum of the potential and \textcolor{black}{the quantities $\Pi_j$ denote higher-order correction terms in the WKB scheme, 
which depend on the overtone number $n$ and on successive derivatives of the 
effective potential with respect to $r_{*}$ computed at its maximum. They extend the original third-order 
WKB formula of \cite{Schutz:1985km} by incorporating higher-order contributions in the 
asymptotic expansion. For instance, in the sixth-order WKB approach, three 
additional correction terms $\Pi_4, \Pi_5$ and $\Pi_6$ are included, as explicitly given in \cite{Konoplya:2003ii}}

 Thus, given the overtone number and the effective potential, it is straightforward to solve Eq.~(\ref{wkb-formula}) for the corresponding $\omega$.

As mentioned, the method has undergone systematic improvements, from the third order to the sixth order, and more recently up to the thirteenth order; see, for instance,\cite{Schutz:1985km,Konoplya:2003ii,Konoplya:2003dd, Matyjasek:2017psv}. For the purposes of this work, the $6^{th}$ WKB expansion is sufficiently precise. As the order of WKB expansion increase, the expressions for functions $\Pi_j$ grows in size, which explicitly expressions can be found in references \cite{Schutz:1985km,Konoplya:2003ii,Matyjasek:2017psv}.

It is important to emphasize that the WKB approximation is, in principle, an asymptotic expansion, and therefore convergence at higher orders is not strictly guaranteed. For this reason, it is necessary to verify that the results obtained at a given WKB order remain consistent, within a numerical accuracy, as the order of the approximation increases. In the specific case considered here, the method exhibits stable and well-behaved convergence up to the sixth order.

Finally, it is worth emphasizing that this method (at different orders of approximation and with certain modifications) has been successfully employed in a wide range of studies (see, for instance \cite{Panotopoulos:2020zbj,Contreras:2025qbx,Yang:2025pmv,Gogoi:2024epx,Lambiase:2023hng,Lambiase:2024lvo,Rincon:2018sgd,Panotopoulos:2020mii,Koch:2025gaw,Pantig:2025eda,Rincon:2025buq,Rincon:2024won,Misyura:2024fho,Balart:2023odm,Rincon:2023hvd,Rincon:2021gwd} and references therein). Also, recent works using this method can be consulted in \cite{ Lutfuoglu:2025bsf, Lutfuoglu:2025hjy, Lutfuoglu:2025hwh, Lutfuoglu:2025ljm, Lutfuoglu:2025qkt, Lutfuoglu:2025blw}.

\subsection{Double-null coordinate integration}

For the AdS geometries, where the \textcolor{black}{WKB} method may not be used considering the shape of the potential assympthotically, we choose another numerical approach the characteristic integration in double null coordinates. In the characteristic integration, we start by considering the perturbation equation (\ref{SLE}) with the field decomposition (\ref{fdc}), such that, the operator $\partial_t$ is casted as $-i\omega$. The double-null coordinates $u$ and $v$ are introduced as the combination of the pair $t$ and $r_*$ as $\partial_t = 2(\partial_u +\partial_v ) $ and $\partial_{r_*} = 2(\partial_v -\partial_u )$. The correspondent field equation turns to 
\be
\label{fedn}
\left(4\frac{\partial^2}{\partial u \partial v} - V(r) \right) \Psi = 0
\ee
which can be easily discretized and integrated (see e. g. \cite{Konoplya:2011qq,Fontana_2021, Mo_2018}) taking physical relevant boundary conditions. The integration recipe is written for the evolution of a "north" point of the field (see e. g. Eq. (22) of  \cite{Fontana_2021}),
\be
\label{ipr}
\Psi_N = \Psi_E + \Psi_W - \Psi_S + h^2 \frac{V_W \Psi_W + V_E \Psi_E}{8}
\ee
and iterated with one piece of Cauchy surface as that of (\ref{pbc}) considering the grid size, $h= \Delta u = \Delta v$, the other represented by the boundary condition in AdS asymptote region. The fundamental issue of such recipe is the definition of the tortoise coordinate as a function of $r$, as long as the potential must be evaluated for each $r_*$. Since in our case, the integral (\ref{tcd1}) can not be solved analytically, we perform numerical approximations \cite{qnmcbtz} to evaluate $V_W$ and $V_E$ of (\ref{ipr}): two expansions in both boundaries of the spacetime, similar to the procedure developed in \cite{qnmcbtz}.

As a first step, we rewrite the variable $r$ as
\be
\label{rwr}
r=\textcolor{black}{a} /y 
\ee
($y_hr_h=\textcolor{black}{a}$) considering a subsequent change in $y$ for a near horizon expansion,
\be
\label{nhe}
y=y_h (1+\varepsilon ),
\ee
$\varepsilon$ small. Then the tortoise coordinate is expressed as
\begin{eqnarray}\nonumber
r_* = -\textcolor{black}{a} \int \frac{1}{y^2\sqrt{1-y^2}-2my^3 - L}dy = \\\nonumber
-\textcolor{black}{a} y_h \int \frac{1}{\underbrace{y_h^2(1+\varepsilon)^2\sqrt{1-y_h^2(1+\varepsilon)^2}}_{s_1(\varepsilon)}  \underbrace{-2my_h^3(1+\varepsilon)^3 - L}_{s_2(\varepsilon)}}d\varepsilon \\
\label{tc1}
\end{eqnarray}
with  $m=M/\textcolor{black}{a}, L=\Lambda \textcolor{black}{a}^2/3$ and $s_j (\varepsilon ) \equiv \sum_{n=0}^{\infty} A_{jn} \varepsilon^n$ the Taylor expansion of both series ($j=1,2$) in $\varepsilon =0$. Now, if we consider 
\be
\label{se3}
s^{-1} =  \frac{1}{s_1+s_2} = \frac{1}{\sum_{n=1}^{\infty} A_{n} \varepsilon^n} \equiv \sum_{n=0}^{\infty} B_{n} \varepsilon^{n-1},  
\ee
the coefficients $A_n$ and $B_n$ are equated as
\begin{eqnarray}\nonumber
\label{cfeq}
B_{n}=-B_0\sum_{j=0}^{n-1}B_{j}A_{1+n-j} \rightarrow n\geq 1, \hspace{0.3cm} B_0=A_1^{-1}\rightarrow n=0, \\
\end{eqnarray}
such that the near horizon expression for the tortoise coordinate is written as
\be
\label{tnh}
r_*^{(nh)} = -\textcolor{black}{a} y_h\left( B_0 \log {\varepsilon} + \sum_{j=0}^{\infty}\frac{B_j\varepsilon^{j+1}}{j+1}\right).
\ee

In the second boundary, $r\rightarrow \infty$, we may consider (\ref{rwr}) with the integral
\be
r_* = -\textcolor{black}{a} \int \frac{1}{\underbrace{y^2\sqrt{1-y^2}}_{s_3(y)}\underbrace{-2my^3 - L}_{s_4(y)}}dy 
\label{tc2}
\ee
in the approach $y\rightarrow 0$, or $s_j(y) = \sum_{n=0}^{\infty} A_{jn} y^n$ and again
\be
\label{ss2}
S^{-1} = \frac{1}{s_3+s_4} = \frac{1}{\sum_{n=0}^{\infty} \mathbb{A}_{n} y^n} \equiv \sum_{n=0}^{\infty} b_{n} y^{n},
\ee
with a similar relation for $\mathbb{A}$ and $b$,
\begin{eqnarray}\nonumber
\label{cfeq2}
b_{n}=-b_0\sum_{j=1}^{n}b_{n-j}\mathbb{A}_{j} \rightarrow n\geq 1, \hspace{0.3cm} b_0=\mathbb{A}_0^{-1}\rightarrow n=0, \\
\end{eqnarray}
that yields the final expression for the tortoise coordinate near $y=0$, 
\be
\label{tni}
r_*^{(ni)} = -\textcolor{black}{a} \sum_{j=0}^{\infty}\frac{b_jy^{j+1}}{j+1}.
\ee

Expressions (\ref{tni}) and (\ref{tnh}) have to be matched at a specific intermediate point (that will depend on the geometry parameters) respecting the convergence of both series in order to assign the tortoise constant (gauge). Once we approximate the value of $r_* (r)$, we then employ a bisection method to invert the relation obtaining $r(r_* )$ and thereafter iterate the recipe prescribed in (\ref{ipr}), obtaining the field evolution. Afterwards, the determination of the quasinormal frequencies is carried out through the Prony method \cite{Konoplya:2011qq}.

\textcolor{black}{To finish the present section we will discuss the convergence of the expansions (\ref{se3})-(\ref{ss2}). As mentioned earlier, Eq. (\ref{tcd1}) is not analytical and in order to obtain the relation $r (r_*)$, we approximate it by (\ref{se3})-(\ref{ss2}). We emphasize that although (\ref{se3}) exhibits good convergence over nearly the entire parameter space of interest, it diverges at the endpoint $r_* =0$ and therefore, it is not a suitable choice for this region. In such case, we must perform the second expansion nearby $r_* =0$ and match both approximations.}

\textcolor{black}{Defining the range of our visible Universe $\mathcal{R}$ such that 
\be
\nonumber
\mathcal{R} \hspace{0.1cm} | \hspace{0.1cm} r_\mathcal{R} \equiv r \in [r_h, \infty ), \hspace{0.2cm} \varepsilon_\mathcal{R} \equiv \varepsilon \in [0,-1), \hspace{0.2cm} y_\mathcal{R} \equiv y \in [y_h, 0) \\
\ee
we can see that $\varepsilon_\mathcal{R}$ and $y_\mathcal{R}$ overlap in their entire range, thus enabling a matching point in-between their extremes. Expansion (\ref{se3}) is most convergent near $\varepsilon =0$ diverging near $\varepsilon \sim -1$; in turn expansion (\ref{ss2}) converges better near $y=0$ and diverges in $y \sim 0$. The existence of a matching point is ensured once we pick a sufficient number of terms in both series. For computational purposes however, once we establish a maximum value for $\Lambda < -1/100M^2$ (so as to prevent the singular point in the expansions, $ \Lambda = 0$ once $y \rightarrow 0$), a physically motivated choice of small $a$ ($a<1$), both series can be truncated picking between 200 to 500 terms, depending on the desired convergence level\footnote{For 500 terms in both series in the scope we established, with a cosmological constant near the singular point, $\Lambda \sim -0.01$, $S^{-1}$ does not deviate more than $10^{-21}$, from (\ref{ss2}) and $s^{-1}$ not more than $10^{-20}$ from (\ref{se3}) in the overlap region. }. We display a few plots of the convergence process of both expansions considering different precisions and number of terms where each series is truncated at appendix \ref{appexpan}. }

\begin{figure*}[ht!]
     \centering
     \begin{subfigure}[b]{0.32\textwidth}
         \centering
         \includegraphics[width=\textwidth]{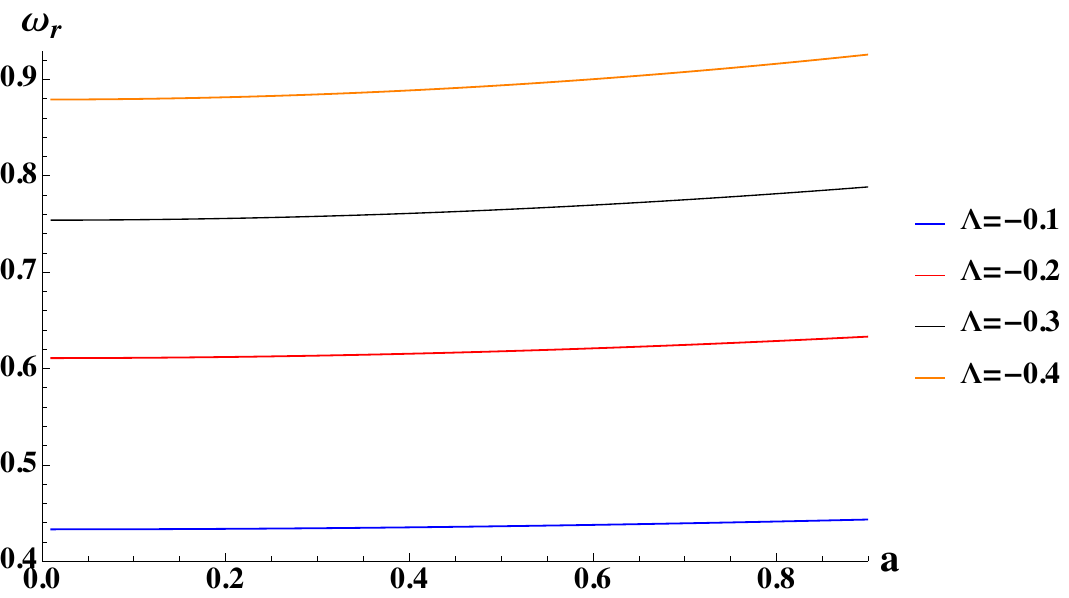}
     \end{subfigure}
     \begin{subfigure}[b]{0.32\textwidth}
         \centering
         \includegraphics[width=\textwidth]{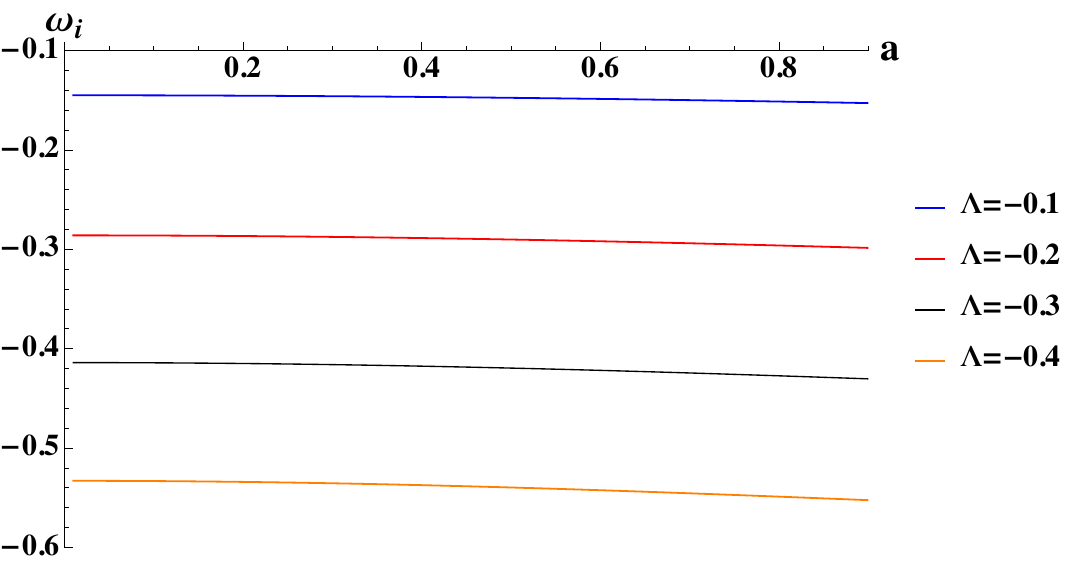}
     \end{subfigure}
     \begin{subfigure}[b]{0.32\textwidth}
         \centering
         \includegraphics[width=\textwidth]{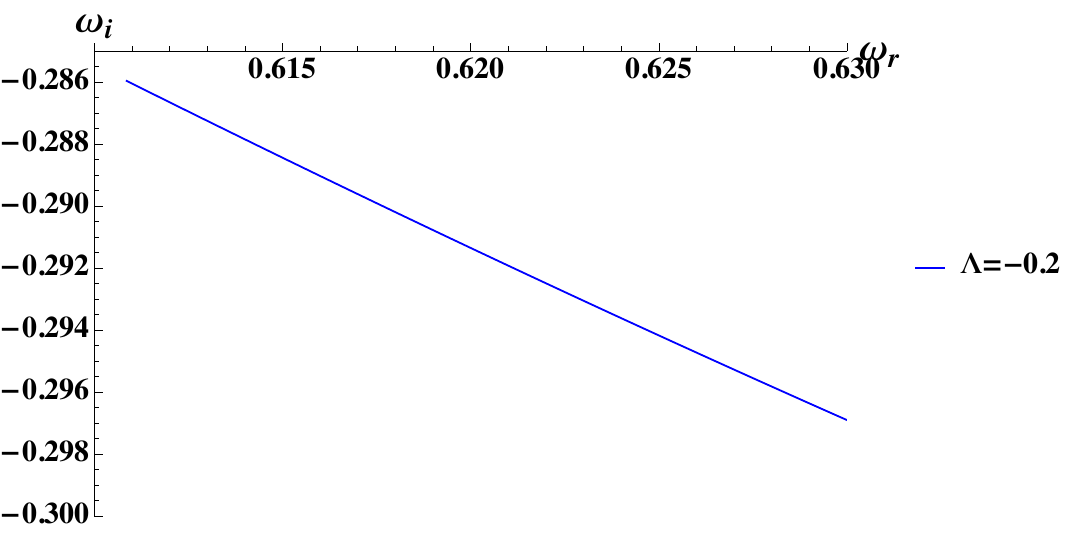}
     \end{subfigure}
      \caption{
    \textcolor{black}{ \footnotesize The quasinormal modes of the quantum AdS black hole submitted to a scalar field perturbation. The geometry and field parameters are $M=1$ and $\mu =\ell =0$.  }}
        \label{fq1}
\end{figure*}

\begin{figure*}[ht!]
     \centering
     \begin{subfigure}[b]{0.49\textwidth}
         \centering
         \includegraphics[width=\textwidth]{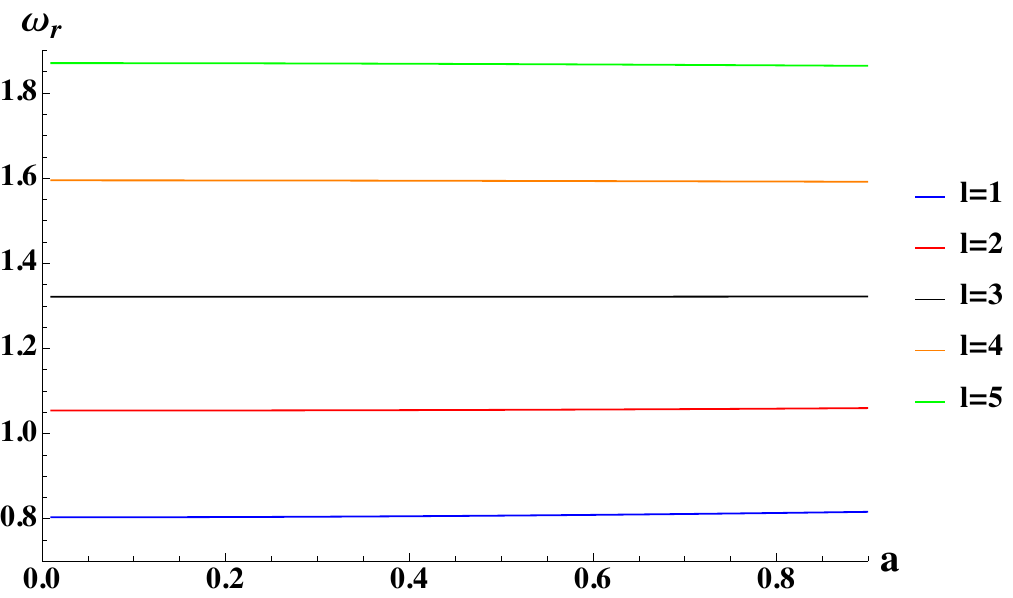}
     \end{subfigure}
     \begin{subfigure}[b]{0.49\textwidth}
         \centering
         \includegraphics[width=\textwidth]{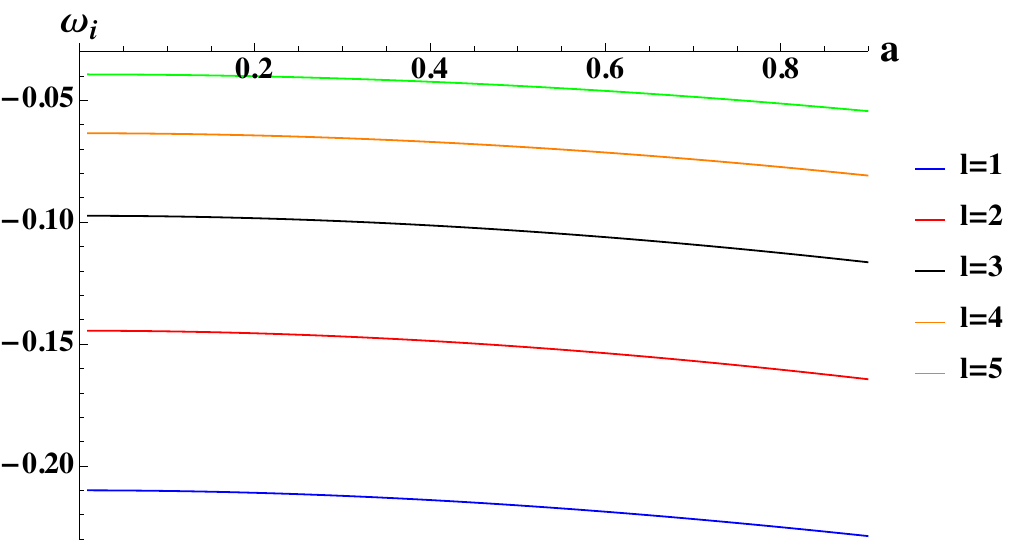}
     \end{subfigure}
       \caption{
    \textcolor{black}{Quasinormal modes for different $\ell$. The geometry parameters read $M=-5\Lambda = 1$.}}
       \label{fq2}
\end{figure*}
\section{Numerical results}
\label{sec:nume}

After briefly reviewing the numerical methods employed to analyze black hole perturbations in the geometries discussed in Section~\ref{sec:back}, we now turn to the presentation of our results. We begin with the AdS-Schwarzschild black hole solution given by Eqs.~(\ref{metric})–(\ref{me1}), and subsequently address the corresponding dS case.

\subsection{ AdS black holes}
\label{ads_rs}

The asymptotic structure of the AdS black holes in the region $r \rightarrow \infty$ determines in the scalar case a divergent potential and in the Dirac case, a constant value. The physical relevant treatment is that of a zero flux o energy in that region that congregates Dirichlet and Neumann boundary conditions \cite{Dappiaggi_2018}. Since we are concerned with the fundamental quasinormal mode, we may adapt the study to the case $\psi = 0$ in both cases \cite{nledbtz, Fontana:2023dix}.

\subsubsection{The scalar field}
The prescription for boundary conditions through the field equation (\ref{SLE}) with the potential (\ref{poten}) can be read as
\be
\label{adssbc}
\psi_{r\rightarrow r_h} &\rightarrow& Ae^{i\omega r_*} + Be^{-i\omega r_*} \\
\label{adssbc2}
\psi_{r\rightarrow \infty} &\rightarrow& 0
\ee
where the physical relevant front wave near $r_h$ is that of amplitude $B$ since no information emerges from the event horizon ($A=0$). The secondary boundary condition comes form the fact that the scalar potential diverges when $r\rightarrow \infty$. Both conditions were briefly presented in section \ref{sp1} with the details we enrich above.  
The results we collected for the scalar field perturbation indicate the absence of instabilities, the perturbations evolving after the initial burst as quasinormal modes.


\textcolor{black}{We begin by validating our numerical methods against results available in the literature, finding excellent agreement in the appropriate limits. In particular, Ref. \cite{Horowitz_2000} reports the fundamental quasinormal mode for $r_h=1$ as $\omega = 2.7982 - 2.6712i$, whereas our implementation yields $\omega = 2.79822 - 2.66986i$. The relative discrepancy between the two results is below $0.1\%$, providing strong evidence for the accuracy, convergence, and reliability of our numerical procedures.}

\textcolor{black}{The agreement improves further for larger black holes. For $r_h=100$, we obtain $\omega = 184.997 - 266.324i$, which differs by only $0.03\%$ from the value reported in Ref. \cite{Horowitz_2000}, $\omega = 184.953 - 266.386i$. These comparisons demonstrate the robustness of our numerical implementation across a broad range of horizon radii.}

Three panels of the behavior of fundamental quasinormal modes are listed in figure (\ref{fq1}).
In these figures we visualize the influence of the extra parameter introduced in the geometry $a$ in the quasinormal frequencies as a mild effect for small $a$ varying just a few percent as a goes to 1. That behavior is very similar for both the imaginary and real parts of the frequencies but going in opposite direction: increasing $a$ increases $\omega_r$ and diminishes $\omega_i$. This relation is nicely linear for small values of $f$ as demonstrated in the third panel of (\ref{fq1}) where a straight line can be observed for $\omega_r$ x $\omega_i$ as we vary $a$, a very peculiar behavior of such theories.


In the figure (\ref{fq2}) we plot the behavior of the field for high angular momenta. Such behavior is very similar to the case $\ell =0$ as we change $a$. The noteworthy feature of those graphics is the pronounced variation of $\omega_i$ as we change $\ell$ behavior not seen in asymptotic plane/dS geometries. 

We finish the AdS black holes section by analyzing the field perturbations of the Dirac field without mass demonstrating it to be stable to first order perturbations. 

\subsubsection{The Dirac field}

The Dirac quasinormal modes follow from the same boundary conditions employed in the scalar case (viz. Eqs. (\ref{adssbc}) and (\ref{adssbc2})) and come as a consequence from  the stability of the spacetime as well. In such case we apply the usual techniques to obtain the spectra for each value of black hole parameter considering a variety of cosmological constants and $a$ in both potentials. 

We deploy the first obtained results in table \ref{tb1} with the oscillations for both potentials $V_{D+}$ and $V_{D-}$. As expected in AdS spacetimes the spectrum is specific for each potential since the behavior at the border is different. The fundamental mode has a much smaller imaginary part for the $V_{D-}$ potential compared to the $V_{D+}$ case. It is however majorly affected by the introduction of the quantum parameter $a$, almost twice as for $V_{D+}$.

\begin{table}[h]
\begin{center}
  \centering
 \caption{The massless Dirac quasinormal modes with $\ell = 1$. The geometry parameters read $M=-10\Lambda =1$.}
\addtolength\tabcolsep{4pt}
    \begin{tabular}{c|c|c}
    \hline 
a	&	$ \omega (V_{D+})$ 	&	$ \omega (V_{D-}) $	\\
\hline \hline
0	&	0.531440 - 0.014128i	&	0.451340 - 0.038334i	\\
0.1	&	0.531417 - 0.014183i	&	0.448531 - 0.038253i	\\
0.2	&	0.531348 - 0.014345i	&	0.440087 - 0.037985i	\\
0.3	&	0.531236 - 0.014615i	&	0.425726 - 0.037449i	\\
0.4	&	0.531085 - 0.014991i	&	0.405186 - 0.036530i	\\
0.5	&	0.530899 - 0.015473i	&	0.378431 - 0.035107i	\\
0.6	&	0.530685 - 0.016060i	&	0.345942 - 0.033110i	\\
0.7	&	0.530449 - 0.016748i	&	0.308917 - 0.030563i	\\
0.8	&	0.530200 - 0.017536i	&	0.269255 - 0.027612i	\\
 \hline  
    \end{tabular}
  \label{tb1}
\end{center}
\end{table}

Interestingly enough, when we increase the cosmological term, the oscillations separate in their qualitative behavior. In table \ref{tb2} we can see the quasinormal spectrum to be purely imaginary for $V_{D-}$ and oscillatory for $V_{D+}$ a dynamical aspect repeated as we keep increasing $\Lambda$.

\begin{table}[h]
\begin{center}
  \centering
 \caption{The massless Dirac quasinormal modes with $\ell = 1$. The geometry parameters read $M=-5\Lambda =1$.}
\addtolength\tabcolsep{4pt}
    \begin{tabular}{c|c|c}
    \hline 
a	&	$ \omega (V_{D+})$ 	&	$ \omega (V_{D-}) $	\\
\hline \hline
0	&	0.727122 - 0.054491i	&	-0.000089286i	\\
0.1	&	0.727134 - 0.054616i	&	-0.000090056i	\\
0.2	&	0.727170 - 0.054987i	&	-0.000092371i	\\
0.3	&	0.727234 - 0.055601i	&	-0.000096319i	\\
0.4	&	0.727331 - 0.056453i	&	-0.000102037i	\\
0.5	&	0.727465 - 0.057534i	&	-0.000109721i	\\
0.6	&	0.727645 - 0.058834i	&	-0.000119634i	\\
0.7	&	0.727878 - 0.060341i	&	-0.000132109i	\\
0.8	&	0.728173 - 0.062042i	&	-0.000147559i	\\
\hline
    \end{tabular}
  \label{tb2}
\end{center}
\end{table}

Also worth of noticing is the fact that the purely imaginary modes are less susceptible to the influence of the quantum parameter $a$ when compared to the oscillatory case.

\textcolor{black}{As a final remark, we emphasize the break of the isospectral behavior of the potentials $V_{D+}$ and $V_{D-}$, as demonstrated in the above tables. The spectra are expected to be the same \cite{ander} for plane waves boundary conditions what is not the case here. Considering an expansion of the potential for expressing the transmission coefficient \cite{Chandrasekhar:1985kt} the spectra of both $V_{D+}$ and $V_{D-}$ will only be the same if this transmission coefficient of a typical scattering theory is the same. In such case, the quasinormal modes (calculated as the poles of such function) are the also same frequencies for both partners.
These is achieved if the following expressions are the same in the boundary \cite{Chandrasekhar:1985kt},
\be
V, \hspace{0.8cm} -\dot{V}, \hspace{0.8cm} -\ddot{V}-V^2, \hspace{0.8cm} -\dddot{V} + 2\dot{V^2}, etc
\ee
in which the dot represents derivative with respect to $r_*$. In our case both $\ddot{V}$ and $\dddot{V}$ are different for $V_{D+}$ and $V_{D-}$, the first one in the pure AdS case and the second enhanced by the quantum factor.}

\textcolor{black}{As a consequence, the spectra of the two potentials are not identical; rather, they are characterized by different families that dominate the spectrum in distinct ways. Finally, it is worthwhile to mention that, even though the first calculation of Dirac quasinormal modes in the Schwarzschild-AdS geometry was restricted to the positive potential \cite{jing1} following the work in the Schwarzschild spacetime \cite{cho1,jingD}, we have recently being shown that the spectrum is not the same for both potentials\footnote{Similar behavior was also reported, e.g., in \cite{deOliveira:2018weu}, although only partial results were obtained in \cite{jingD2} based on a misleading assumption of isospectrality. } \cite{herdeiroAdS}.}

\subsection{dS black holes}

In black holes with a cosmological horizon, the boundary condition is determined by the plane wave evolving in that region since no divergences are found in the potential there. In what follows we discuss our results obtained through the \textcolor{black}{WKB} method as above described. 

\begin{figure*}[htbp!]
    \centering
    \includegraphics[width=0.46\textwidth]{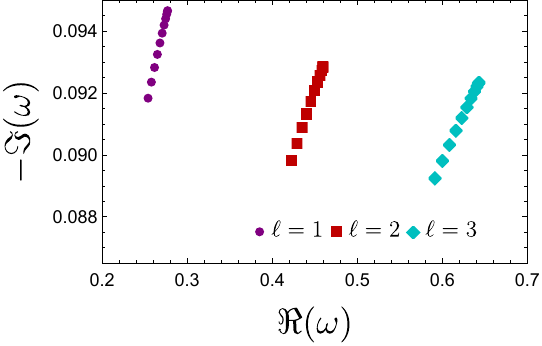}
    \includegraphics[width=0.46\textwidth]{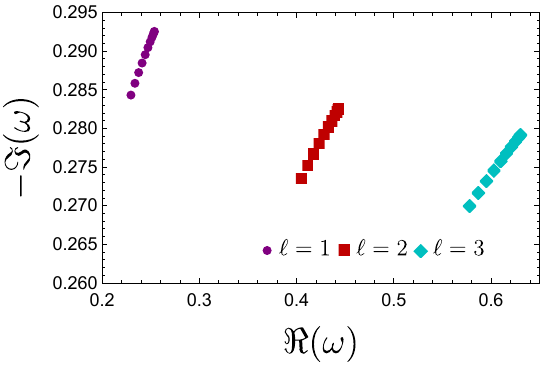}
   \caption{ \footnotesize
   		{\bf Left panel:} Fundamental quasinormal mode ($n = 0$) of massless scalar perturbations computed using the WKB approach with $M = 1$ and $\Lambda = 0.01$, varying the quantum parameter $a$ in the range $0.0 \leq a \leq 1.0$ for three different values of $\ell$ (see figure for details). 
   		{\bf Right panel:} First overtone ($n = 1$) under the same conditions, with $M = 1$ and $\Lambda = 0.01$, showing the dependence on $a$ for the same set of $\ell$ values.
   	}
   \label{fig:Scalar-modes}
\end{figure*}

\begin{figure*}[htbp]
\begin{center}
\includegraphics[width=0.46\linewidth]{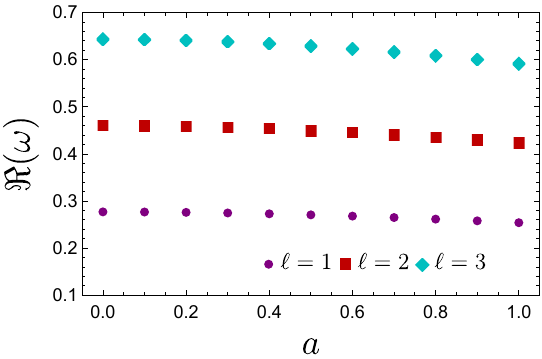} 
\includegraphics[width=0.46\linewidth]{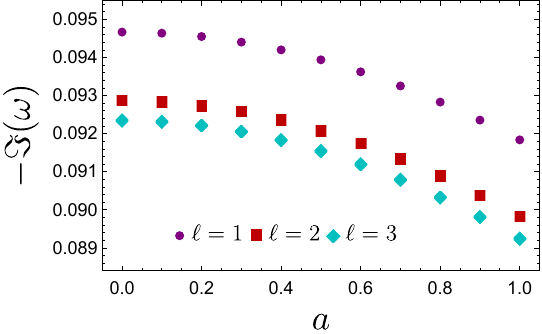}
\\
\vspace{0.5cm}
\includegraphics[width=0.46\linewidth]{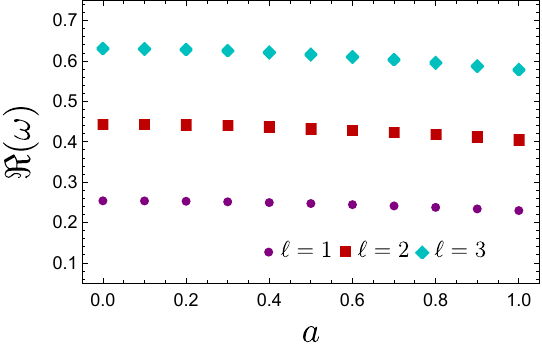} 
\includegraphics[width=0.46\linewidth]{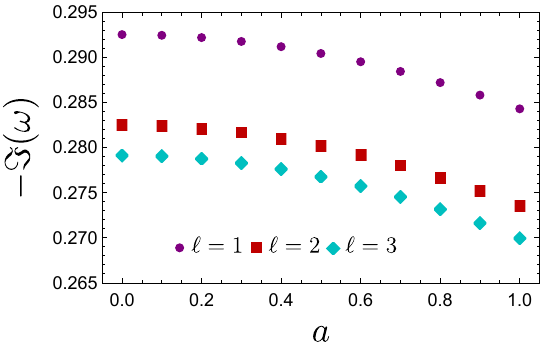}
\caption{\footnotesize Real and imaginary parts of the scalar quasinormal frequencies, $\mathfrak{R}(\omega)$ and $-\mathfrak{I}(\omega)$, as functions of the quantum parameter $a$. 
		{\bf Top panels:} Fundamental mode ($n=0$) for $M=1$ and several values of $\ell$. 
		{\bf Bottom panels:} First overtone ($n=1$) under the same conditions. 
		Left and right plots correspond, respectively, to the real and imaginary parts of the frequencies.
	}
\label{fig:real_imag_var_a}
\end{center}
\end{figure*}

\subsubsection{The Scalar Field}

The calculated quasinormal frequencies are listed in Table~\ref{table:First set} and \ref{table:second_set}, while the overall behavior is illustrated in Figures~(\ref{fig:Scalar-modes}) and (\ref{fig:real_imag_var_a}). In these figures, $\mathfrak{R}(\omega)$ and $\mathfrak{I}(\omega)$ denote the real and imaginary parts of the frequency, respectively. Our numerical findings indicate that all modes remain stable under massless scalar perturbations, as evidenced by the negative sign of the imaginary component of $\omega$.

Figure~\ref{fig:Scalar-modes} displays the relation between the real part $\mathfrak{R}(\omega)$ and the negative of the imaginary part $\mathfrak{I}(\omega)$ of the quasinormal frequencies for different values of the multipole number $\ell$ and varying quantum parameter $a$. In the left panel, corresponding to the fundamental mode ($n=0$), the data show an almost linear relation between $\mathfrak{R}(\omega)$ and $-\mathfrak{I}(\omega)$ for each $\ell$. As the quantum parameter $a$ increases, both the real and imaginary parts of the frequency decrease, indicating lower oscillation frequencies and slower damping rates. The right panel shows the same qualitative behavior for the first overtone ($n=1$), although the magnitude of the imaginary part is slightly more sensitive to variations in $a$. In both cases, the curves remain well separated for distinct $\ell$, showing that higher angular momentum values correspond to larger real frequencies and smaller damping.



 In Figure~\ref{fig:real_imag_var_a} is showed the dependence of the real and imaginary parts of the scalar quasinormal frequencies on the quantum parameter $a$. The top panels correspond to the fundamental mode ($n=0$) for different multipole numbers $\ell$, while the bottom panels display the first overtone ($n=1$) under the same conditions. In both cases, $\mathfrak{R}(\omega)$ and $-\mathfrak{I}(\omega)$ decrease monotonically as $a$ increases, indicating that quantum corrections tend to lower both the oscillation frequency and the damping rate of the perturbations. The effect is slightly more pronounced for higher $\ell$ and for the first overtone, reflecting the sensitivity of excited modes to changes in the background geometry. Overall, the results confirm that the parameter $a$ diminish the dynamical response of the system, leading to longer lived and less energetic scalar oscillations.
 
\textcolor{black}{As a validation test, we considered the limit $a=0$, for which the geometry reduces to the Schwarzschild-de Sitter spacetime. The scalar quasinormal frequencies obtained with our WKB implementation are in agreement with the results obtained in \cite{Zhidenko:2003wq}, as shown in Table~\ref{tab:scalar_sds_comparison}.}

\begin{table}[h]
\begin{center}
  \centering
 \caption{The scalar quasinormal modes with $\ell = 2$ and overtone number $n=1$. The geometry parameters read $M=1$ and $a=0$ for comparison with \cite{Zhidenko:2003wq}.}
\addtolength\tabcolsep{4pt}
{\color{black}
    \begin{tabular}{c|c|c}
    \hline 
$\Lambda $    &	$ \omega $ 	&	$ \omega$ from \cite{Zhidenko:2003wq}	\\
\hline \hline
0.00	&	0.463847 -  0.295625 I	&	0.46385 - 0.29563 I	\\
0.02	&	0.420841 -  0.268616 I	&	0.42084 - 0.26862 I	\\
0.09	&	0.200977 -  0.127941 I	&	0.20098 - 0.12794 I	\\
 \hline  
    \end{tabular}
    }
  \label{tab:scalar_sds_comparison}
\end{center}
\end{table}


\subsubsection{The Dirac field}

Similar to the massless scalar case, we summarize our main results for massless Dirac perturbations in the presence of a positive cosmological constant through one table and several illustrative figures. Tables~\ref{table:First_set_dirac} and~\ref{table:second_set_dirac} list the quasinormal (QN) frequencies computed for $M = 1$, $\Lambda = 0.01$, and different combinations of the parameters $\{n, a, \xi\}$.

\begin{figure*}[htbp]
    \centering
    \includegraphics[width=0.46\textwidth]{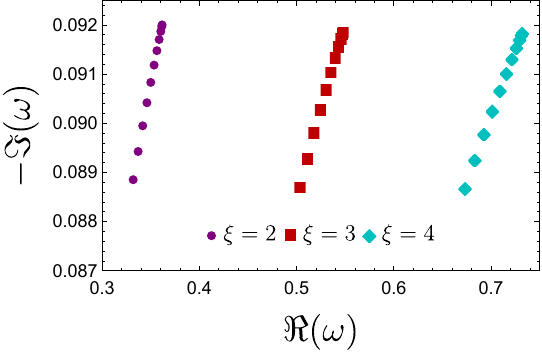}
    \includegraphics[width=0.46\textwidth]{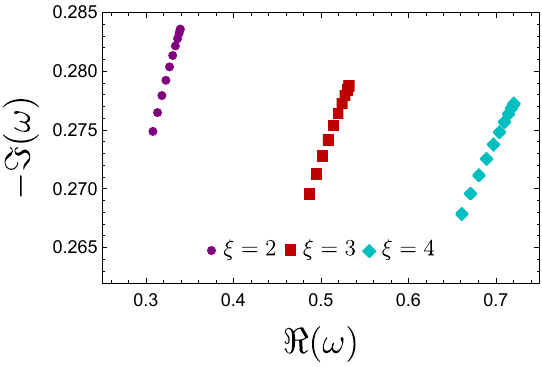}
    \caption{\footnotesize
    		{\bf Left panel:} Fundamental quasinormal mode ($n = 0$) of Dirac perturbations  with $M = 1$ and $\Lambda = 0.01$, varying the quantum parameter $a$ in the range $0.0 \leq a \leq 1.0$ for three different values of $\xi$ (see figure for details). 
    		{\bf Right panel:} First overtone ($n = 1$) under the same conditions, with $M = 1$ and $\Lambda = 0.01$, showing the dependence on $a$ for the same set of $\xi$ values.
    	}
    \label{fig:Dirac-modes}
\end{figure*}

In the panels do Figure~\ref{fig:Dirac-modes} are displayed the relation between  $\mathfrak{R}(\omega)$ and  $-\mathfrak{I}(\omega)$ of the quasinormal. The qualitative behavior closely resembles that observed for scalar perturbations in Figure~\ref{fig:Scalar-modes}. For each value of the constant $\xi$, the results reveal an approximately linear correlation between $\mathfrak{R}(\omega)$ and $-\mathfrak{I}(\omega)$ as the quantum parameter $a$ increases. Both quantities decrease monotonically with $a$, indicating as in the scalar case that quantum corrections reduce the oscillation frequency and the damping rate of the Dirac modes. The curves for distinct $\xi$ remain well separated, showing that higher angular indices correspond to larger real frequencies and smaller damping, in full agreement with the behavior found in the scalar case.

\begin{figure*} [htbp]
\begin{center}
\includegraphics[width=0.46\linewidth]{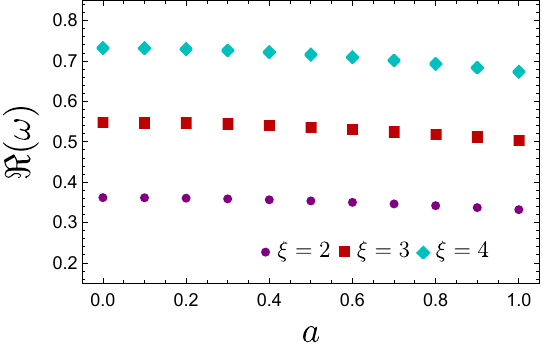} 
\includegraphics[width=0.46\linewidth]{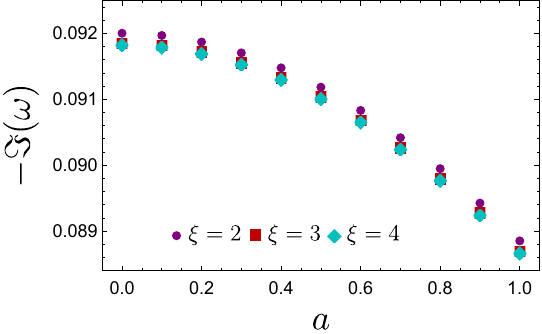}
\\
\vspace{0.5cm}
\includegraphics[width=0.46\linewidth]{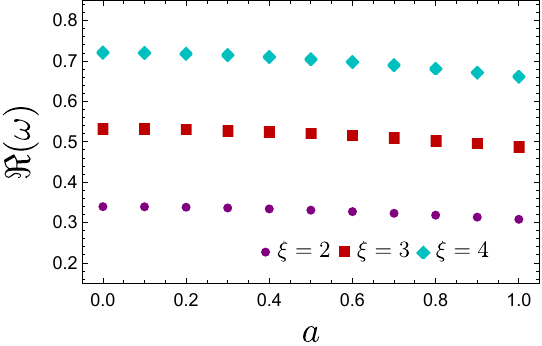} 
\includegraphics[width=0.46\linewidth]{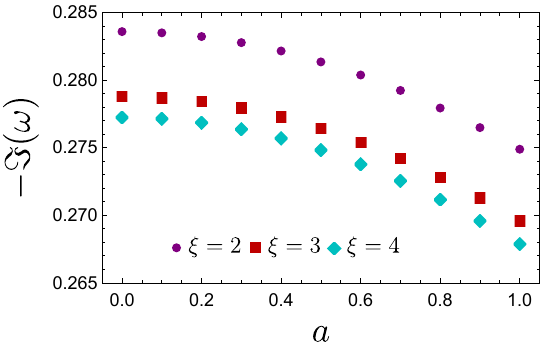}
\caption{\footnotesize
		Real and imaginary parts of the Dirac quasinormal frequencies, $\mathfrak{R}(\omega)$ and $-\mathfrak{I}(\omega)$, as functions of the quantum parameter $a$. 
		{\bf Top panels:} Fundamental mode ($n=0$) for $M=1$ and several values of $\xi$. 
		{\bf Bottom panels:} First overtone ($n=1$) under the same conditions. 
	}
\label{Fig4-1}
\end{center}
\end{figure*}

Figure~\ref{Fig4-1} illustrates how the real and imaginary components of the Dirac quasinormal frequencies vary with the quantum parameter $a$. The upper panels correspond to the fundamental mode ($n=0$), while the lower ones show the first overtone ($n=1$), each plotted for different values of the constant $\xi$. As $a$ increases, both $\mathfrak{R}(\omega)$ and $-\mathfrak{I}(\omega)$ gradually decrease, indicating that the quantum corrections tend to suppress the oscillatory behavior and extend the decay time of the perturbations. This attenuation effect is slightly stronger for the first overtone, revealing its greater sensitivity to the near-horizon modifications induced by the parameter $a$. The overall result closely follows that observed for scalar perturbations in Figure~\ref{fig:real_imag_var_a}, confirming that the influence of $a$ on the quasinormal spectrum is qualitatively similar in both the scalar and spinorial sectors.

\textcolor{black}{We also verified the Schwarzschild - de Sitter limit for Dirac perturbations by setting $a=0$. The frequencies obtained in our calculations are consistent with those available in the literature \cite{Zhidenko:2003wq}, as summarized in Table~\ref{tab:dirac_sds_comparison}, confirming the correctness of our numerical implementation.}

\begin{table}[h]
\begin{center}
  \centering
 \caption{The massless Dirac quasinormal modes with $\ell = 1$. The geometry parameters read $M=1$ and $a=0$ for comparison with classical results.}
\addtolength\tabcolsep{4pt}
{\color{black}
    \begin{tabular}{c|c|c}
    \hline 
$\Lambda $    &	$ \omega (V_{D+})$ 	&	$ \omega (V_{D-}) $	\\
\hline \hline
0.000	&	0.378627 - 0.0965425 I	&	0.379712 - 0.0964590 I	\\
0.020	&	0.343756 - 0.0872558 I	&	0.344641 - 0.0871877 I	\\
0.040	&	0.304515 - 0.0769786 I	&	0.305206 - 0.0769279 I	\\
0.060	&	0.258891 - 0.0651916 I	&	0.259393 - 0.0651754 I	\\
0.080	&	0.202289 - 0.0496666 I	&	0.202910 - 0.0509241 I	\\
0.090	&	0.169279 - 0.0497585 I	&	0.167177 - 0.0412133 I	\\
0.100	&	0.123453 - 0.0376956 I	&	0.123180 - 0.0363292 I	\\
 \hline  
    \end{tabular}
    }
  \label{tab:dirac_sds_comparison}
\end{center}
\end{table}

\newpage

\section{Discussion and final remarks}
\label{sec:disc}
In this article, we have investigated the quasinormal modes of a black hole with quantum corrections in four dimensions using two alternative methods. We have studied how this black hole solution responds to two types of perturbations, scalar and spin 1/2 (Dirac) particles. Our results have been summarized in a collection of tables and figures for different values of the parameters involved. In particular, we have explored in detail the field evolution for the case of positive and negative cosmological constants.

First, for AdS black holes perturbed by a test scalar field, our results show no instabilities, meaning that the perturbations evolve after the initial burst as towers of quasinormal modes. For the Dirac field, the isospectrality typically present in GR is broken (i.e., $\omega ( V_{D+} ) \neq \omega ( V_{D-} )$). Notably, as the cosmological constant increases, the quasinormal spectrum becomes purely imaginary for $V_{D-}$ and oscillatory for $V_{D+}$, indicating the presence of different families of oscillations dominating the spectrum in each case.

Finally, we turn our attention to the dS case. 
In light of the negative sign of the imaginary component of $\omega$, our numerical findings confirm that all modes remain stable under massless scalar perturbations.
For Dirac perturbations, the black hole responds in the same way as in the scalar case, i.e. the system is stable against Dirac perturbations. Additionally, as $a$ increases, both the real and imaginary parts gradually decrease, indicating
that quantum corrections suppress oscillatory behavior and extend decay time of the perturbations.

\textcolor{black}{
It is also interesting to compare our results with previous studies of quasinormal modes in black holes arising from higher-curvature or higher-derivative corrections to general relativity. Although a direct quantitative comparison is not straightforward, since those works consider different gravitational theories and perturbative sectors, a qualitative comparison can still be done.}

\textcolor{black}{
 In the de Sitter case, increasing the quantum parameter $a$ decreases both the real part of the quasinormal frequency and the the imaginary part, leading to oscillations with lower frequency and longer decay times. A qualitatively similar behavior has been found for test scalar perturbations of Einstein-dilaton-Gauss-Bonnet black holes, where the Gauss-Bonnet coupling modifies the oscillation frequency and damping rate relative to the Schwarzschild case \cite{Konoplya:2019}.
}

\textcolor{black}{
For the AdS case with scalar perturbations, however, we observe a different situation: increasing the parameter $a$ increases the real part of the frequency while the imaginary part becomes is diminished. This behavior contrasts with that found in studies of eikonal quasinormal modes of string-corrected black holes, where higher-derivative corrections typically lead to a increasing th both real and imaginary parts of quasnormal frequencies \cite{Moura:2021PLB}. 
}

\textcolor{black}{
More generally, studies of quasinormal modes in higher-derivative or string-inspired black hole geometries \cite{Blazquez-Salcedo:2016enn,Cano:2020,Moura:2021JHEP,Moura:2023,Konoplya:2023} consistently show that deviations from general relativity modify the quasinormal spectrum, although the precise results depends strongly on the  gravitational theory, the asymptotic structure of the spacetime, and the type of perturbation considered (test scalar field, Dirac, electromagnetic or gravitational). In this sense, the behavior found here for the quantum parameter $a$ fits naturally within the broader picture in which corrections beyond Einstein gravity leave characteristic signatures in the ringdown spectrum of black holes.
}

Further lines of investigation include similar geometries with charge and cosmological constant in which superradiance phenomena may take action if charged fields are considered.

\section*{Acknowledgments}

A.~R. would like to express his gratitude to Silesian University in Opava, Czech Republic, for their financial support.
J. Oliveira is funded by {\it Conselho Nacional de Desenvolvimento Científico e Tecnológico} (CNPq) under Grant No. $405749/2023-6$ and {\it Fundação de Amparo à Pesquisa do Estado de Mato Grosso} (FAPEMAT) under Grant No. 010/2025. R. D. B. Fontana acknowledges {\it Conselho Nacional de Desenvolvimento Científico e Tecnológico} (CNPq) under Grants No. $309304/2026-1$, $407868/2025-9$ and $405749/2023-6$.
\appendix
\section{Tables with the spectrum of scalar QNMs for de Sitter case }
\begin{table*}[htbp]
	\centering
	\caption{ \footnotesize
		\textcolor{black}{
			Quasinormal frequencies for massless scalar perturbations (varying $\ell$, $n$ and $a$) fixing $M=1,\Lambda=0.01$ for the model considered in this work using the WKB approximation.}}
	\resizebox{1.6\columnwidth}{!}{
		{
			\begin{tabular}{c|c|ccc} 
				$a$  &$n$  &  $\ell=1$ & $\ell=2$ & $\ell=3$  
				\\ 
				\hline
				0.0 & 0 & 0.276926\, -0.094668 i & 0.459636\, -0.092860 i & 0.642975\, -0.092343 i \\
				0.0 & 1 &    $\times \times$     & 0.443015\, -0.282479 i & 0.630447\, -0.279146 i \\
				0.0 & 2 &    $\times \times$     &    $\times \times$     & 0.606844\, -0.472194 i \\
				\hline
				0.1 & 0 & 0.276664\, -0.094639 i & 0.459211\, -0.092828 i & 0.642383\, -0.092310 i \\
				0.1 & 1 &    $\times \times$     & 0.442578\, -0.282386 i & 0.629847\, -0.279050 i \\
				0.1 & 2 &    $\times \times$     &    $\times \times$     & 0.606226\, -0.472039 i \\
				\hline
				0.2 & 0 & 0.275884\, -0.094551 i & 0.457941\, -0.092733 i & 0.640618\, -0.092213 i \\
				0.2 & 1 &    $\times \times$     & 0.441275\, -0.282106 i & 0.628056\, -0.278763 i \\
				0.2 & 2 &    $\times \times$     &    $\times \times$     & 0.604384\, -0.471573 i \\
				\hline
				0.3 & 0 & 0.274601\, -0.094404 i & 0.455850\, -0.092574 i & 0.637710\, -0.092052 i \\
				0.3 & 1 &    $\times \times$     & 0.439129\, -0.281642 i & 0.625106\, -0.278285 i \\
				0.3 & 2 &    $\times \times$     &    $\times \times$     & 0.601351\, -0.470799 i \\
				\hline
				0.4 & 0 & 0.272832\, -0.094200 i & 0.452972\, -0.092354 i & 0.633709\, -0.091827 i \\
				0.4 & 1 &    $\times \times$     & 0.436178\, -0.280994 i & 0.621049\, -0.277619 i \\
				0.4 & 2 &    $\times \times$     &    $\times \times$     & 0.597179\, -0.469721 i \\
				\hline
				0.5 & 0 & 0.270609\, -0.093939 i & 0.449355\, -0.092073 i & 0.628681\, -0.091541 i \\
				0.5 & 1 &    $\times \times$     & 0.432470\, -0.280168 i & 0.615951\, -0.276769 i \\
				0.5 & 2 &    $\times \times$     &    $\times \times$     & 0.591941\, -0.468343 i \\
				\hline
				0.6 & 0 & 0.267966\, -0.093623 i & 0.445055\, -0.091732 i & 0.622704\, -0.091194 i \\
				0.6 & 1 &    $\times \times$     & 0.428065\, -0.279165 i & 0.609893\, -0.275740 i \\
				0.6 & 2 &    $\times \times$     &    $\times \times$     & 0.585720\, -0.466672 i \\
				\hline
				0.7 & 0 & 0.264939\, -0.093252 i & 0.440135\, -0.091334 i & 0.615867\, -0.090788 i \\
				0.7 & 1 &    $\times \times$     & 0.423028\, -0.277993 i & 0.602967\, -0.274535 i \\
				0.7 & 2 &    $\times \times$     &    $\times \times$     & 0.578613\, -0.464715 i \\
				\hline
				0.8 & 0 & 0.261571\, -0.092830 i & 0.434664\, -0.090880 i & 0.608264\, -0.090326 i \\
				0.8 & 1 &    $\times \times$     & 0.417433\, -0.276655 i & 0.595268\, -0.273162 i \\
				0.8 & 2 &    $\times \times$     &    $\times \times$     & 0.570720\, -0.462482 i \\
				\hline
				0.9 & 0 & 0.257903\, -0.092357 i & 0.428711\, -0.090373 i & 0.599992\, -0.089810 i \\
				0.9 & 1 &    $\times \times$     & 0.411350\, -0.275160 i & 0.586898\, -0.271627 i \\
				0.9 & 2 &    $\times \times$     &    $\times \times$     & 0.562147\, -0.459983 i \\
				\hline
				1.0 & 0 & 0.253977\, -0.091837 i & 0.422345\, -0.089817 i & 0.591148\, -0.089242 i \\
				1.0 & 1 &    $\times \times$     & 0.404853\, -0.273515 i & 0.577955\, -0.269939 i \\
				1.0 & 2 &    $\times \times$     &    $\times \times$     & 0.552998\, -0.457231 i \\
			\end{tabular} 
			\label{table:First set}
		}
	}
\end{table*}


\begin{table*}[htbp!]
	\centering
	\caption{\footnotesize
		\textcolor{black}{
			Quasinormal frequencies for massless scalar perturbations (varying $\ell$, $n$ and $a$) fixing $M=1,\Lambda=0.01$ for the model considered in this work using the WKB approximation.
		}
	}
	{
		\resizebox{1.6\columnwidth}{!}
		{\begin{tabular}{c|c|cccc} 
				$a$  &$n$  &  $\ell=4$ & $\ell=5$ & $\ell=6$ & $\ell=7$ 
				\\ 
				\hline
				0.0 & 0 & 0.826449\, -0.092127 i & 1.009970\, -0.092017 i & 1.193510\, -0.091954 i & 1.377070\, -0.091914 i \\
				0.0 & 1 & 0.816484\, -0.277695 i & 1.001720\, -0.276941 i & 1.186490\, -0.276503 i & 1.370960\, -0.276225 i \\
				0.0 & 2 & 0.797249\, -0.467172 i & 0.985612\, -0.464526 i & 1.172670\, -0.462972 i & 1.358880\, -0.461987 i \\
				0.0 & 3 & 0.770199\, -0.662959 i & 0.962438\, -0.656468 i & 1.152550\, -0.652613 i & 1.341160\, -0.650152 i \\
				\hline
				0.1 & 0 & 0.825690\, -0.092094 i & 1.009040\, -0.091984 i & 1.192420\, -0.091921 i & 1.375810\, -0.091881 i \\
				0.1 & 1 & 0.815718\, -0.277597 i & 1.000790\, -0.276843 i & 1.185390\, -0.276404 i & 1.369690\, -0.276127 i \\
				0.1 & 2 & 0.796469\, -0.467013 i & 0.984669\, -0.464364 i & 1.171570\, -0.462810 i & 1.357610\, -0.461824 i \\
				0.1 & 3 & 0.769399\, -0.662742 i & 0.961477\, -0.656247 i & 1.151430\, -0.652389 i & 1.339870\, -0.649926 i \\
				\hline
				0.2 & 0 & 0.823427\, -0.091997 i & 1.006280\, -0.091886 i & 1.189160\, -0.091822 i & 1.372050\, -0.091782 i \\
				0.2 & 1 & 0.813434\, -0.277306 i & 0.998013\, -0.276550 i & 1.182120\, -0.276110 i & 1.365920\, -0.275832 i \\
				0.2 & 2 & 0.794144\, -0.466536 i & 0.981856\, -0.463881 i & 1.168260\, -0.462324 i & 1.353810\, -0.461335 i \\
				0.2 & 3 & 0.767013\, -0.662094 i & 0.958613\, -0.655583 i & 1.148080\, -0.651717 i & 1.336040\, -0.649248 i \\
				\hline
				0.3 & 0 & 0.819699\, -0.091834 i & 1.001730\, -0.091723 i & 1.183790\, -0.091659 i & 1.365850\, -0.091619 i \\
				0.3 & 1 & 0.809673\, -0.276822 i & 0.993436\, -0.276064 i & 1.176720\, -0.275622 i & 1.359700\, -0.275342 i \\
				0.3 & 2 & 0.790316\, -0.465743 i & 0.977223\, -0.463079 i & 1.162820\, -0.461516 i & 1.347550\, -0.460524 i \\
				0.3 & 3 & 0.763085\, -0.661016 i & 0.953897\, -0.654481 i & 1.142560\, -0.650600 i & 1.329720\, -0.648123 i \\
				\hline
				0.4 & 0 & 0.814570\, -0.091608 i & 0.995473\, -0.091496 i & 1.176400\, -0.091431 i & 1.357330\, -0.091391 i \\
				0.4 & 1 & 0.804499\, -0.276149 i & 0.987139\, -0.275386 i & 1.169300\, -0.274941 i & 1.351150\, -0.274660 i \\
				0.4 & 2 & 0.785051\, -0.464639 i & 0.970851\, -0.461961 i & 1.155330\, -0.460390 i & 1.338950\, -0.459394 i \\
				0.4 & 3 & 0.757684\, -0.659514 i & 0.947412\, -0.652944 i & 1.134980\, -0.649044 i & 1.321030\, -0.646555 i \\
				\hline
				0.5 & 0 & 0.808124\, -0.091319 i & 0.987607\, -0.091206 i & 1.167110\, -0.091141 i & 1.346620\, -0.091099 i \\
				0.5 & 1 & 0.797997\, -0.275289 i & 0.979226\, -0.274520 i & 1.159970\, -0.274073 i & 1.340410\, -0.273790 i \\
				0.5 & 2 & 0.778437\, -0.463228 i & 0.962845\, -0.460534 i & 1.145920\, -0.458953 i & 1.328130\, -0.457950 i \\
				0.5 & 3 & 0.750902\, -0.657594 i & 0.939266\, -0.650981 i & 1.125450\, -0.647056 i & 1.310110\, -0.644552 i \\
				\hline
				0.6 & 0 & 0.800462\, -0.090969 i & 0.978257\, -0.090855 i & 1.156070\, -0.090789 i & 1.333890\, -0.090747 i \\
				0.6 & 1 & 0.790271\, -0.274246 i & 0.969823\, -0.273471 i & 1.148880\, -0.273020 i & 1.327640\, -0.272735 i \\
				0.6 & 2 & 0.770581\, -0.461518 i & 0.953334\, -0.458804 i & 1.134750\, -0.457211 i & 1.315280\, -0.456201 i \\
				0.6 & 3 & 0.742851\, -0.655264 i & 0.929593\, -0.648600 i & 1.114140\, -0.644647 i & 1.297140\, -0.642124 i \\
				\hline
				0.7 & 0 & 0.791698\, -0.090561 i & 0.967562\, -0.090444 i & 1.143440\, -0.090377 i & 1.319320\, -0.090335 i \\
				0.7 & 1 & 0.781436\, -0.273027 i & 0.959069\, -0.272245 i & 1.136210\, -0.271789 i & 1.313030\, -0.271501 i \\
				0.7 & 2 & 0.761601\, -0.459517 i & 0.942461\, -0.456779 i & 1.121970\, -0.455174 i & 1.300590\, -0.454156 i \\
				0.7 & 3 & 0.733654\, -0.652535 i & 0.918540\, -0.645812 i & 1.101200\, -0.641826 i & 1.282310\, -0.639284 i \\
				\hline
				0.8 & 0 & 0.781953\, -0.090095 i & 0.955670\, -0.089977 i & 1.129400\, -0.089908 i & 1.303130\, -0.089866 i \\
				0.8 & 1 & 0.771615\, -0.271638 i & 0.947115\, -0.270848 i & 1.122110\, -0.270387 i & 1.296790\, -0.270096 i \\
				0.8 & 2 & 0.751626\, -0.457234 i & 0.930379\, -0.454471 i & 1.107760\, -0.452851 i & 1.284250\, -0.451824 i \\
				0.8 & 3 & 0.723445\, -0.649419 i & 0.906264\, -0.642632 i & 1.086840\, -0.638610 i & 1.265830\, -0.636045 i \\
				\hline
				0.9 & 0 & 0.771352\, -0.089574 i & 0.942734\, -0.089454 i & 1.114120\, -0.089385 i & 1.285520\, -0.089341 i \\
				0.9 & 1 & 0.760935\, -0.270086 i & 0.934113\, -0.269286 i & 1.106780\, -0.268820 i & 1.279130\, -0.268525 i \\
				0.9 & 2 & 0.740785\, -0.454681 i & 0.917244\, -0.451891 i & 1.092320\, -0.450255 i & 1.266490\, -0.449218 i \\
				0.9 & 3 & 0.712361\, -0.645930 i & 0.892928\, -0.639073 i & 1.071220\, -0.635012 i & 1.247920\, -0.632423 i \\
				\hline
				1.0 & 0 & 0.760018\, -0.089002 i & 0.928904\, -0.088879 i & 1.097790\, -0.088809 i & 1.266690\, -0.088764 i \\
				1.0 & 1 & 0.749522\, -0.268378 i & 0.920217\, -0.267568 i & 1.090400\, -0.267096 i & 1.260250\, -0.266797 i \\
				1.0 & 2 & 0.729209\, -0.451870 i & 0.903215\, -0.449051 i & 1.075820\, -0.447398 i & 1.247520\, -0.446351 i \\
				1.0 & 3 & 0.700537\, -0.642084 i & 0.878694\, -0.635153 i & 1.054550\, -0.631050 i & 1.228800\, -0.628436 i \\
				\hline
			\end{tabular} 
			\label{table:second_set}
		}
	}
\end{table*}

\newpage

\section{Tables with the spectrum of Dirac QNMs}

\begin{table*}[t]
	\centering
	\caption{\footnotesize
		\textcolor{black}{
			Quasinormal frequencies for massless Dirac perturbations (varying $\xi$, $n$ and $a$) fixing $M=1,\Lambda=0.01$ for the model considered in this work using the WKB approximation.
		}
	}
	{
		\resizebox{1.4\columnwidth}{!}
		{\begin{tabular}{c|c|ccc} 
				$a$  &$n$  &  $\xi=2$ & $\xi=3$ & $\xi=4$  
				\\ 
				\hline
				0.0 & 0 & 0.361653\, -0.092002 i & 0.547556\, -0.091847 i & 0.732072\, -0.091819 i \\
				0.0 & 1 & 0.339344\, -0.283588 i & 0.532166\, -0.278796 i & 0.720494\, -0.277239 i \\
				0.0 & 2 &    $\times \times$     & 0.506291\, -0.471884 i & 0.699699\, -0.466718 i \\
				0.0 & 3 &    $\times \times$     &    $\times \times$     & 0.672252\, -0.660239 i \\
				\hline
				0.1 & 0 & 0.361313\, -0.091968 i & 0.547051\, -0.091814 i & 0.731400\, -0.091786 i \\
				0.1 & 1 & 0.338981\, -0.283495 i & 0.531648\, -0.278698 i & 0.719813\, -0.277141 i \\
				0.1 & 2 &    $\times \times$     & 0.505751\, -0.471725 i & 0.699003\, -0.466558 i \\
				0.1 & 3 &    $\times \times$     &    $\times \times$     & 0.671535\, -0.660017 i \\
				\hline
				0.2 & 0 & 0.360298\, -0.091868 i & 0.545544\, -0.091716 i & 0.729394\, -0.091687 i \\
				0.2 & 1 & 0.337905\, -0.283222 i & 0.530104\, -0.278411 i & 0.717782\, -0.276848 i \\
				0.2 & 2 &    $\times \times$     & 0.504150\, -0.471260 i & 0.696926\, -0.466077 i \\
				0.2 & 3 &    $\times \times$     &    $\times \times$     & 0.669395\, -0.659351 i \\
				\hline
				0.3 & 0 & 0.358627\, -0.091704 i & 0.543062\, -0.091551 i & 0.726091\, -0.091523 i \\
				0.3 & 1 & 0.336136\, -0.282773 i & 0.527562\, -0.277931 i & 0.714436\, -0.276361 i \\
				0.3 & 2 &    $\times \times$     & 0.501508\, -0.470481 i & 0.693506\, -0.465278 i \\
				0.3 & 3 &    $\times \times$     &    $\times \times$     & 0.665873\, -0.658244 i \\
				\hline
				0.4 & 0 & 0.356329\, -0.091477 i & 0.539647\, -0.091323 i & 0.721547\, -0.091294 i \\
				0.4 & 1 & 0.333710\, -0.282154 i & 0.524066\, -0.277264 i & 0.709835\, -0.275683 i \\
				0.4 & 2 &    $\times \times$     & 0.497879\, -0.469396 i & 0.688803\, -0.464165 i \\
				0.4 & 3 &    $\times \times$     &    $\times \times$     & 0.661032\, -0.656702 i \\
				\hline
				0.5 & 0 & 0.353440\, -0.091184 i & 0.535357\, -0.091031 i & 0.715837\, -0.091002 i \\
				0.5 & 1 & 0.330651\, -0.281345 i & 0.519675\, -0.276411 i & 0.704054\, -0.274818 i \\
				0.5 & 2 &    $\times \times$     & 0.493324\, -0.468010 i & 0.682899\, -0.462743 i \\
				0.5 & 3 &    $\times \times$     &    $\times \times$     & 0.654958\, -0.654731 i \\
				\hline
				0.6 & 0 & 0.350009\, -0.090831 i & 0.530259\, -0.090677 i & 0.709051\, -0.090649 i \\
				0.6 & 1 & 0.327030\, -0.280376 i & 0.514460\, -0.275378 i & 0.697186\, -0.273770 i \\
				0.6 & 2 &    $\times \times$     & 0.487919\, -0.466328 i & 0.675889\, -0.461020 i \\
				0.6 & 3 &    $\times \times$     &    $\times \times$     & 0.647752\, -0.652339 i \\
				\hline
				0.7 & 0 & 0.346086\, -0.090418 i & 0.524429\, -0.090265 i & 0.701290\, -0.090235 i \\
				0.7 & 1 & 0.322889\, -0.279234 i & 0.508501\, -0.274170 i & 0.689336\, -0.272544 i \\
				0.7 & 2 &    $\times \times$     & 0.481750\, -0.464358 i & 0.667880\, -0.459003 i \\
				0.7 & 3 &    $\times \times$     &    $\times \times$     & 0.639528\, -0.649538 i \\
				\hline
				0.8 & 0 & 0.341726\, -0.089948 i & 0.517948\, -0.089795 i & 0.692662\, -0.089765 i \\
				0.8 & 1 & 0.318299\, -0.277935 i & 0.501882\, -0.272794 i & 0.680612\, -0.271146 i \\
				0.8 & 2 &    $\times \times$     & 0.474911\, -0.462111 i & 0.658989\, -0.456702 i \\
				0.8 & 3 &    $\times \times$     &    $\times \times$     & 0.630407\, -0.646340 i \\
				\hline
				0.9 & 0 & 0.336987\, -0.089426 i & 0.510901\, -0.089270 i & 0.683278\, -0.089239 i \\
				0.9 & 1 & 0.313323\, -0.276486 i & 0.494692\, -0.271255 i & 0.671129\, -0.269585 i \\
				0.9 & 2 &    $\times \times$     & 0.467492\, -0.459594 i & 0.649334\, -0.454128 i \\
				0.9 & 3 &    $\times \times$     &    $\times \times$     & 0.620515\, -0.642758 i \\
				\hline
				1.0 & 0 & 0.331924\, -0.088852 i & 0.503370\, -0.088694 i & 0.673248\, -0.088661 i \\
				1.0 & 1 & 0.308017\, -0.274886 i & 0.487018\, -0.269562 i & 0.660999\, -0.267867 i \\
				1.0 & 2 &    $\times \times$     & 0.459589\, -0.456820 i & 0.639033\, -0.451293 i \\
				1.0 & 3 &    $\times \times$     &    $\times \times$     & 0.609978\, -0.638811 i \\
				\hline
			\end{tabular} 
			\label{table:First_set_dirac}
		}
	}
\end{table*}

\begin{table*}[t]
	\centering
	\caption{\footnotesize
		\textcolor{black}{
			Quasinormal frequencies for massless Dirac perturbations (varying $\xi$, $n$ and $a$) fixing $M=1,\Lambda=0.01$ for the model considered in this work using the WKB approximation.
		}
	}
	{
		\resizebox{1.6\columnwidth}{!}
		{\begin{tabular}{c|c|cccc} 
				$a$  &$n$  &  $\xi=5$ & $\xi=6$ & $\xi=7$ & $\xi=8$ 
				\\ 
				\hline
				0.0 & 0 & 0.916158\, -0.091808 i & 1.100058\, -0.091803 i & 1.283861\, -0.091800 i & 1.467607\, -0.091798 i \\
				0.0 & 1 & 0.906891\, -0.276550 i & 1.092334\, -0.276185 i & 1.277240\, -0.275968 i & 1.461812\, -0.275829 i \\
				0.0 & 2 & 0.889651\, -0.464109 i & 1.077665\, -0.462616 i & 1.264499\, -0.461686 i & 1.450565\, -0.461068 i \\
				0.0 & 3 & 0.866171\, -0.654985 i & 1.057224\, -0.651711 i & 1.246454\, -0.649548 i & 1.434448\, -0.648053 i \\
				\hline
				0.1 & 0 & 0.915318\, -0.091775 i & 1.099050\, -0.091770 i & 1.282686\, -0.091767 i & 1.466264\, -0.091765 i \\
				0.1 & 1 & 0.906044\, -0.276452 i & 1.091321\, -0.276086 i & 1.276059\, -0.275869 i & 1.460465\, -0.275730 i \\
				0.1 & 2 & 0.888791\, -0.463946 i & 1.076641\, -0.462453 i & 1.263310\, -0.461522 i & 1.449210\, -0.460904 i \\
				0.1 & 3 & 0.865294\, -0.654760 i & 1.056186\, -0.651484 i & 1.245252\, -0.649320 i & 1.433081\, -0.647825 i \\
				\hline
				0.2 & 0 & 0.912812\, -0.091676 i & 1.096045\, -0.091671 i & 1.279180\, -0.091668 i & 1.462257\, -0.091666 i \\
				0.2 & 1 & 0.903519\, -0.276157 i & 1.088299\, -0.275791 i & 1.272540\, -0.275573 i & 1.456446\, -0.275433 i \\
				0.2 & 2 & 0.886229\, -0.463461 i & 1.073588\, -0.461964 i & 1.259763\, -0.461031 i & 1.445168\, -0.460413 i \\
				0.2 & 3 & 0.862682\, -0.654085 i & 1.053090\, -0.650805 i & 1.241669\, -0.648638 i & 1.429006\, -0.647140 i \\
				\hline
				0.3 & 0 & 0.908686\, -0.091512 i & 1.091094\, -0.091507 i & 1.273405\, -0.091504 i & 1.455658\, -0.091502 i \\
				0.3 & 1 & 0.899360\, -0.275668 i & 1.083322\, -0.275300 i & 1.266743\, -0.275082 i & 1.449828\, -0.274942 i \\
				0.3 & 2 & 0.882010\, -0.462653 i & 1.068561\, -0.461152 i & 1.253923\, -0.460217 i & 1.438511\, -0.459596 i \\
				0.3 & 3 & 0.858380\, -0.652966 i & 1.047992\, -0.649677 i & 1.235767\, -0.647505 i & 1.422295\, -0.646004 i \\
				\hline
				0.4 & 0 & 0.903009\, -0.091283 i & 1.084284\, -0.091278 i & 1.265461\, -0.091275 i & 1.446580\, -0.091273 i \\
				0.4 & 1 & 0.893638\, -0.274986 i & 1.076475\, -0.274617 i & 1.258767\, -0.274398 i & 1.440722\, -0.274257 i \\
				0.4 & 2 & 0.876208\, -0.461529 i & 1.061646\, -0.460021 i & 1.245889\, -0.459083 i & 1.429354\, -0.458460 i \\
				0.4 & 3 & 0.852465\, -0.651404 i & 1.040982\, -0.648105 i & 1.227650\, -0.645926 i & 1.413064\, -0.644420 i \\
				\hline
				0.5 & 0 & 0.895875\, -0.090991 i & 1.075725\, -0.090986 i & 1.255477\, -0.090983 i & 1.435171\, -0.090981 i \\
				0.5 & 1 & 0.886450\, -0.274116 i & 1.067872\, -0.273745 i & 1.248745\, -0.273524 i & 1.429280\, -0.273382 i \\
				0.5 & 2 & 0.868920\, -0.460092 i & 1.052959\, -0.458578 i & 1.235795\, -0.457634 i & 1.417848\, -0.457008 i \\
				0.5 & 3 & 0.845040\, -0.649408 i & 1.032179\, -0.646097 i & 1.217454\, -0.643909 i & 1.401468\, -0.642398 i \\
				\hline
				0.6 & 0 & 0.887397\, -0.090637 i & 1.065554\, -0.090632 i & 1.243611\, -0.090629 i & 1.421611\, -0.090627 i \\
				0.6 & 1 & 0.877909\, -0.273062 i & 1.057649\, -0.272688 i & 1.236836\, -0.272466 i & 1.415682\, -0.272323 i \\
				0.6 & 2 & 0.860264\, -0.458351 i & 1.042639\, -0.456828 i & 1.223802\, -0.455879 i & 1.404177\, -0.455249 i \\
				0.6 & 3 & 0.836225\, -0.646989 i & 1.021725\, -0.643662 i & 1.205344\, -0.641465 i & 1.387692\, -0.639946 i \\
				\hline
				0.7 & 0 & 0.877700\, -0.090224 i & 1.053920\, -0.090218 i & 1.230040\, -0.090215 i & 1.406101\, -0.090213 i \\
				0.7 & 1 & 0.868143\, -0.271829 i & 1.045958\, -0.271452 i & 1.223216\, -0.271228 i & 1.400131\, -0.271084 i \\
				0.7 & 2 & 0.850371\, -0.456315 i & 1.030842\, -0.454781 i & 1.210091\, -0.453826 i & 1.388545\, -0.453193 i \\
				0.7 & 3 & 0.826158\, -0.644157 i & 1.009778\, -0.640813 i & 1.191503\, -0.638605 i & 1.371945\, -0.637079 i \\
				\hline
				0.8 & 0 & 0.866920\, -0.089753 i & 1.040985\, -0.089747 i & 1.214951\, -0.089744 i & 1.388858\, -0.089742 i \\
				0.8 & 1 & 0.857288\, -0.270424 i & 1.032963\, -0.270043 i & 1.208076\, -0.269818 i & 1.382842\, -0.269673 i \\
				0.8 & 2 & 0.839381\, -0.453992 i & 1.017734\, -0.452447 i & 1.194852\, -0.451485 i & 1.371170\, -0.450848 i \\
				0.8 & 3 & 0.814982\, -0.640925 i & 0.996513\, -0.637563 i & 1.176127\, -0.635343 i & 1.354448\, -0.633809 i \\
				\hline
				0.9 & 0 & 0.855194\, -0.089227 i & 1.026916\, -0.089220 i & 1.198538\, -0.089217 i & 1.370100\, -0.089215 i \\
				0.9 & 1 & 0.845486\, -0.268854 i & 1.018831\, -0.268469 i & 1.191609\, -0.268242 i & 1.364039\, -0.268095 i \\
				0.9 & 2 & 0.827440\, -0.451395 i & 1.003485\, -0.449838 i & 1.178285\, -0.448870 i & 1.352278\, -0.448228 i \\
				0.9 & 3 & 0.802850\, -0.637309 i & 0.982102\, -0.633928 i & 1.159418\, -0.631696 i & 1.335430\, -0.630154 i \\
				\hline
				1.0 & 0 & 0.842660\, -0.088648 i & 1.011877\, -0.088641 i & 1.180993\, -0.088638 i & 1.350050\, -0.088636 i \\
				1.0 & 1 & 0.832875\, -0.267127 i & 1.003729\, -0.266738 i & 1.174011\, -0.266508 i & 1.343941\, -0.266360 i \\
				1.0 & 2 & 0.814691\, -0.448536 i & 0.988267\, -0.446967 i & 1.160586\, -0.445991 i & 1.332092\, -0.445345 i \\
				1.0 & 3 & 0.789910\, -0.633325 i & 0.966722\, -0.629925 i & 1.141578\, -0.627681 i & 1.315118\, -0.626130 i \\
				\hline
			\end{tabular} 
			\label{table:second_set_dirac}
		}
	}
\end{table*}

\clearpage
\onecolumngrid

\section{Numerical convergence of the tortoise coordinate expansions}
\label{appexpan}
\begin{figure}[h]
    \centering
    \includegraphics[width=0.4\textwidth]{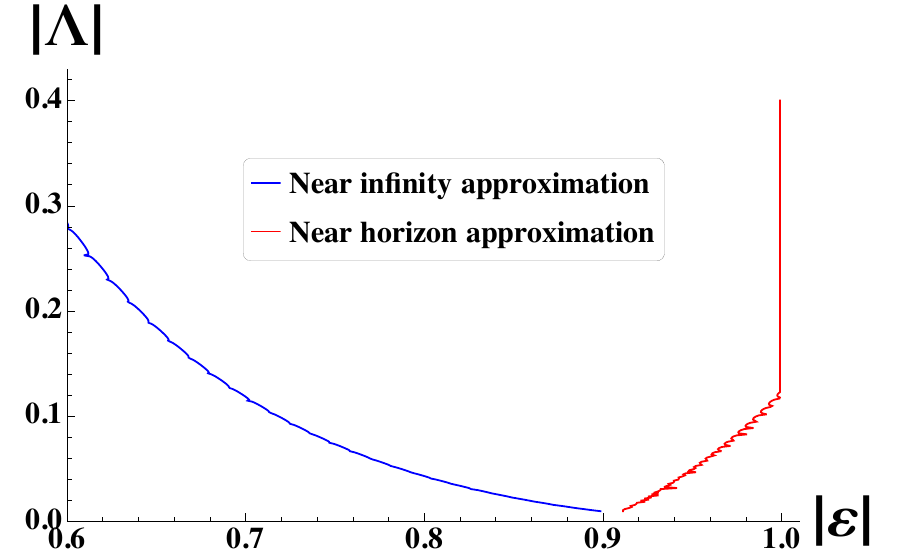}
    \includegraphics[width=0.4\textwidth]{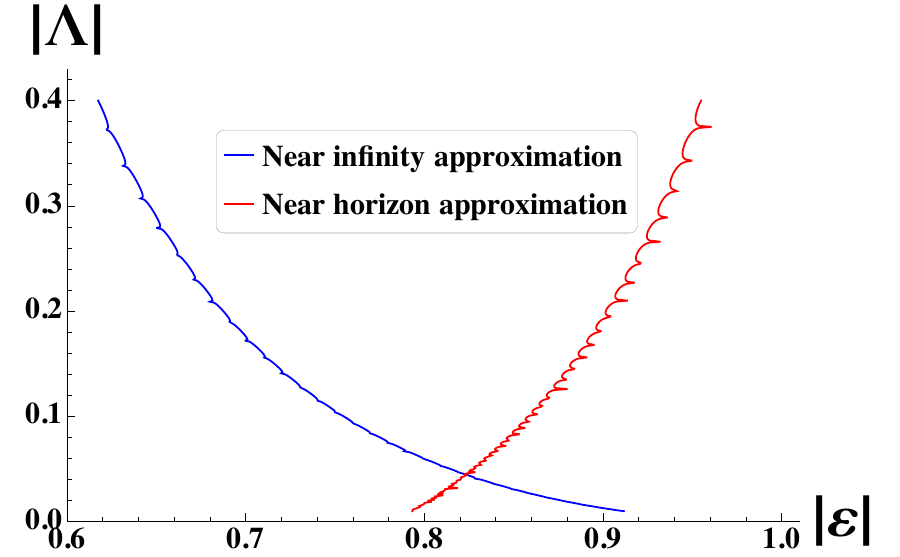}
    \caption{ The convergence of a near horizon (nh) and near infinity (ni) expansions of the tortoise vector, considering expansions with 500 terms and precision of $10^{-20}$ (left) or $10^{-50}$ (right). The threshold for ni is represented by blue curves, being the series more convergent than the required precision in the right regions of the curve. In nh case, the series is more convergent to the left of the red curve. In both panels we can see the overlap region in-between both lines.}
\label{fappc1}
\end{figure}

\begin{figure}
    \centering
    \includegraphics[width=0.4\textwidth]{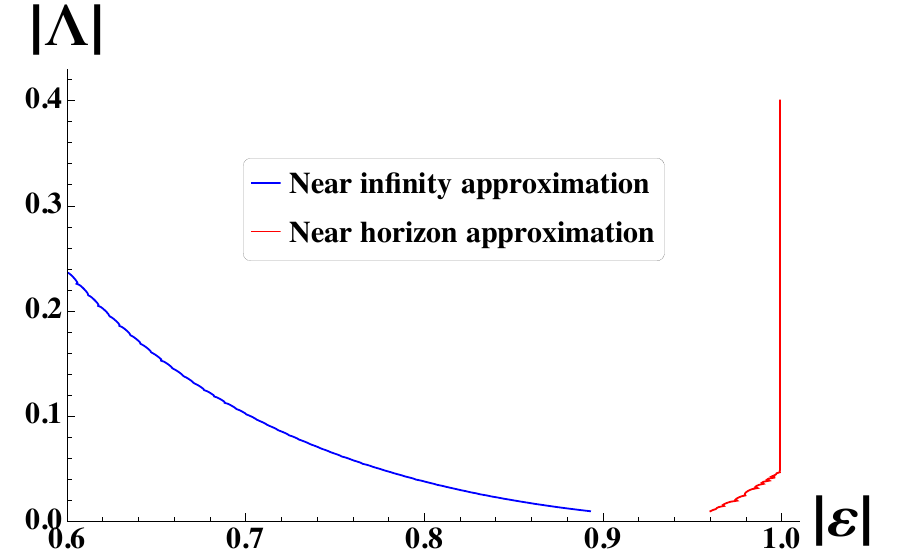}
    \includegraphics[width=0.4\textwidth]{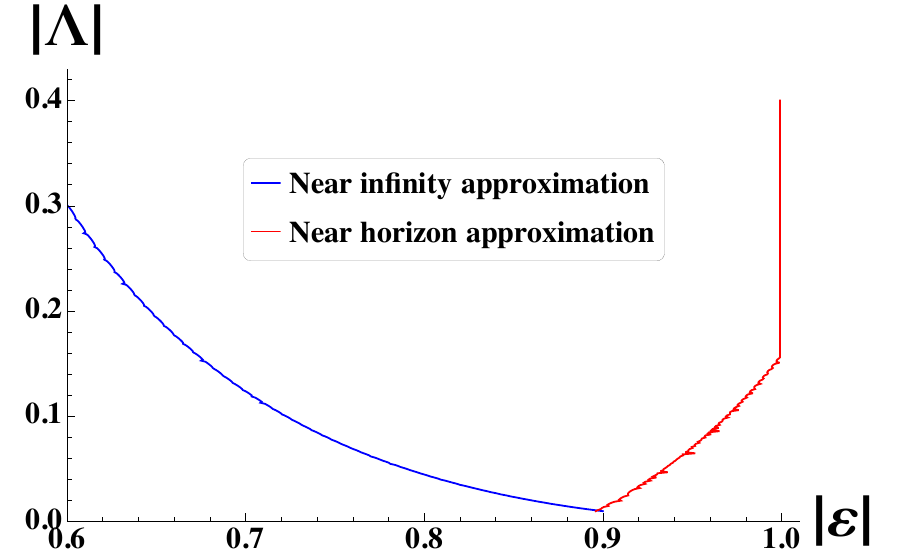}
    \caption{ The convergence of a near horizon (nh) and near infinity (ni) expansions of the tortoise vector, considering expansions with 1000 terms and precision of $10^{-20}$ (left) or $10^{-50}$ (right). The threshold for ni is represented by blue curves, being the series more convergent than the required precision in the right regions of the curve. In nh case, the series is more convergent to the left of the red curve. In both panels we can see the overlap region in-between both lines.}
\label{fappc2}
\end{figure}

In figures \ref{fappc1} and \ref{fappc2} we display different panels with the convergence of both series (\ref{ss2}) and (\ref{se3}). The approximation is tested with different convergence requirement and truncation (number of terms) for both expansions.

\twocolumngrid

\newpage
\bibliographystyle{utphys}
\bibliography{Library}

\end{document}